\documentclass[fleqn,usenatbib]{mnras}

\usepackage{newtxtext,newtxmath}
\usepackage[T1]{fontenc}
\usepackage{ae,aecompl}

\usepackage{graphicx}	% Including figure files
\usepackage{amsmath}	% Advanced maths commands
\usepackage{amssymb,multirow}	% Extra maths symbols
\usepackage{gensymb}
\usepackage{bm}

\newcommand{\pd}[2]{\frac{\partial #1}{\partial #2}}
\newcommand{\dd}[2]{\frac{\mathrm{d} #1}{\mathrm{d} #2}}
\newcommand{\id}{\mathop{}\!\mathrm{d}}
\newcommand{\im}{\mathrm{i}}
\newcommand{\me}{\mathrm{e}}
\newcommand{\cs}{c_\mathrm{s}^2}
\newcommand{\ab}{\alpha_\mathrm{b}}
\newcommand{\rin}{r_{\mathrm{in}}}
\newcommand{\rout}{r_{\mathrm{out}}}
\newcommand{\Oin}{\Omega_{\mathrm{in}}}

\newcommand{\ee}{\left| E\right|^2}

\newcommand{\Mp}{M_{\rm p}}
\newcommand{\Md}{M_{\rm d}}
\newcommand{\Ms}{M_{*}}
\newcommand{\Msun}{M_{\odot}}

\newcommand{\Ed}{E_{\rm d}}
\newcommand{\Ep}{E_{\rm p}}
\newcommand{\ap}{a_{\rm p}}
\newcommand{\ep}{e_{\rm p}}
\newcommand{\Op}{\Omega_{\rm p}}

\newcommand{\Jd}{J_{\rm d}}
\newcommand{\Jp}{J_{\rm p}}
\newcommand{\od}{\omega_{\rm d,p}}
\newcommand{\op}{\omega_{\rm p,d}}
\newcommand{\vd}{\nu_{\rm d,p}}
\newcommand{\vp}{\nu_{\rm p,d}}
\newcommand{\Mj}{M_{\rm J}}

\newcommand{\twopi}{\,2\pi r \,\id r}

\title[Disc-planet eccentricities]{A Simplified Model for the Secular Dynamics of Eccentric Discs and Applications to Planet-Disc Interactions}
\author[Teyssandier \& Lai]{Jean Teyssandier\thanks{E-mail: jt553@cornell.edu} and Dong Lai\\ %$^{1}$
Department of Astronomy, Cornell Center for Astrophysics and Planetary Science, Cornell University, Ithaca, NY 14853, USA
}

\date{Accepted XXX. Received YYY; in original form ZZZ}

\pubyear{2019}

\begin{document}
\label{firstpage}
\pagerange{\pageref{firstpage}--\pageref{lastpage}}
\maketitle

\begin{abstract}
We develop a simplified model for studying the long-term evolution of giant planets in protoplanetary discs. The model accounts for the eccentricity evolution of the planets and the dynamics of eccentric discs under the influences of secular planet-disc interactions and internal disc pressure, self-gravity and viscosity. Adopting the ansatz that the disc precesses coherently with aligned apsides, the eccentricity evolution equations of the planet-disc system reduce to a set of linearized ODEs, which allows for fast computation of the evolution of planet-disc eccentricities over long timescales.  Applying our model to ``giant planet + external disc'' systems, we are able to reproduce and explain the secular behaviours found in previously published hydrodynamical simulations. We re-examine the possibility of eccentricity excitation (due to secular resonance) of multiple planets embedded in a dispersing disc, and find that taking into account the dynamics of eccentric discs can significantly affect the evolution of the planets' eccentricities.
\end{abstract}


\section{Introduction}

The origin of the eccentricity distribution of extrasolar giant planets remains poorly understood. While gravitational interactions between planets and with external stellar companions can play important roles, planet-disc interactions provide the seeds or "floors" for the initial  eccentricities of planets formed in protoplanetary discs.

Early work on the subject was developed in the context of satellites and	 planetary rings, by \citet{gt80,gt81}. When a satellite is massive enough to open a gap in the ring, eccentric Lindblad resonances (ELR) lead to eccentricity excitation, while eccentric corotation resonance (ECR) lead (in general) to eccentricity damping. \citet{gt80} argued that ECRs overcome ELRs by a small amount, leading to an overall damping of eccentricity. %
\citet{gs03} and \citet{ol03} considered giant planet-disc interactions and showed that ECRs undergo non-linear saturation as the eccentricity increases, and become ineffective at damping eccentricities. This suggests that a gap-opening planet may undergo eccentricity growth if its initial eccentricity is larger than a (small) critical value.

More recently, \citet{to16} developed a more complete picture of eccentricity evolution during disc-planet interactions. Their work went beyond that of \citet{gs03} by taking into account the back reaction of the planet on the disc, and deriving a wave equation that describes the evolution of eccentricity in the disc under the effect of pressure, viscosity, self-gravity and external forcing. They presented a careful treatment of the width and torque density distribution of both ELRs and ECRs, and showed that growth of eccentricity is possible even when ECRs are not saturated.

Along with these theoretical developments, advances in computational simulations have also allowed for this problem to be tackled numerically. \citet{papaloizou01} showed that the eccentricity of a massive body (larger than 20 Jupiter masses) is excited due to interactions with a disc, which itself also becomes eccentric. 
Various authors \citep{kd06,regaly10,to17} have simulated the eccentricity dynamics of discs with  gap-opening planets on fixed circular orbits. One general finding is that an initially circular disc can become eccentric when the planet-to-star mass ratio is larger than $\sim 0.003$. Above this threshold, \citet{to17} showed that the growth rate and precession rate of the disc eccentricity agree well with the linear theory of \citet{to16}. Recently, \citet{muley19} recovered this result by allowing the planet to evolve freely in the disc and grow in mass as it accretes material from its surroundings.

Recent observations are providing further motivation for understanding the growth of eccentricity during planet-disc interactions. 
For example, radial velocity measurements of CI Tau, a young (~2 Myrs) T Tauri star with a disc, suggested the presence of a giant planet (~10 Jupiter masses) on a 9 day orbit with eccentricity of 0.3 \citep{johnskrull16}.
\cite{rosotti} and \citet{ragusa18} carried out two two-dimensional hydrodynamical simulations of planet-disc interactions with different disc masses. In one of their simulations	 the eccentricity of the planet grew over very long timescales ($>10^5$ planetary orbits) to reach $\sim 0.12$, and was still growing by the end of the simulation. This work showed that planet-disc interaction could explain the eccentricity of this putative planet.

Planet-disc interactions have been recently invoked by \citet{petrovich19} to explain the orbital architecture of the Kepler-419 system, where two giant planets were detected with high eccentricities and anti-aligned arguments of pericentres. The authors proposed that, as the disc exterior to the planets disperses (e.g. due to photoevaporation), the precession rates induced by its gravitational potential on the planets change, and the planets eventually cross a secular resonance, during which eccentricity excitation and apsidal anti-alignment occur. A key assumption of this work is that the disc remains circular and is not affected by the planets. 
This assumption cannot be justified as the disc inevitably develops eccentricity. As we show in this paper (Section \ref{sec:kepler419}), including the dynamics of eccentric discs significantly changes the effectiveness of the mechanism of \citet{petrovich19}.

Hydrodynamical simulations of planet-disc interactions \citep[such as those by][]{ragusa18} are time-consuming, particularly when long-term evolution (over $10^5$ orbits) is required to capture the  ``correct'' answer. The linear theory of eccentric discs \citet{to16} reduces the problem to one spatial dimension,   but still involves complicated integral-differential equations. The main goal of this paper is to formulate a simple model of planet-disc interactions that can efficiently follow the disc eccentricity evolution.
In essence, we develop a ``reduced'' version of the \citet{to16} theory by adopting a ``trial wavefunction'' ansatz for the disc eccentricity profile which otherwise must be solved in a self-consistent way. This ansatz is necessarily approximate, but allows us to transform the complicated integral-differential equations for the evolution of disc-planet eccentricities into simple ordinary differential equations. For many situations, these ODEs capture the essential physics of the secular interactions between the planet and disc, and can be easily integrated over long timescales.
	
The plan of the paper is as follows. In Section \ref{sec:eq_pd} we introduce our "reduced" model for the dynamics of eccentric discs and planet-disc interactions. In Section \ref{sec:eq_disc} we show how internal processes in the disc (pressure, self-gravity and viscosity) can be included in this model; we derive scaling laws for the relevant precession/damping frequencies of this model in Section \ref{sec:toy}. In Section \ref{sec:ragusa} we show that our model agrees well with the outcome of hydrodynamical simulations, and we apply it to the case of Kepler-419 in Section \ref{sec:kepler419}. Finally we discuss the limitations of our model in Section \ref{sec:discussion}, and conclude in Section \ref{sec:conclusion}.

\section{Evolution equations for an eccentric disc--planet system}
\label{sec:eq_pd}

\subsection{Notations}

To a first approximation, fluid elements in an eccentric disc follow elliptical Keplerian orbits. The eccentricity $e$ and argument of pericentre $\varpi$ vary smoothly with the orbital distance from the star. For small gradients in $e$ and $\varpi$, the orbits are nested without intersections.

For small (linear) eccentricities, it is convenient to present the evolutionary equations in terms of the complex eccentricity $E(r,t)=e\,\me^{\im \varpi}$. Since $|E|=e(r,t)$ and $\arg{(E)}=\varpi(r,t)$, this variable conveniently describes both the shape and the orientation of the elliptical orbits. With similar notations, the planet has a complex eccentricity $\Ep(t) =\ep\,\me^{\im \varpi_{\rm p}} $. We denote by $\ap$, $\Op$ and $\Mp$ the semi-major axis, orbital frequency and mass of the planet. In addition we denote by $\Md$ the total mass of the disc.

\subsection{Secular evolution of a planet-disc system}
\label{sec:sec}

In this section we only consider the long-term (secular) interaction between a planet and a disc. The disc is represented by a continuum of nested eccentric rings. Including only gravitational interactions between the planet and disc, the equations of motion are as follows \citep{to16}:
\begin{align}
\Sigma r^2\Omega \pd{E}{t}&=\im G\Mp\Sigma\left[K_1(r,\ap) E-K_2(r,\ap)\Ep\right],  \label{eq:epd}\\
\Mp\ap^2\Op\dd{\Ep}{t}&=\int\im G\Mp\Sigma\left[K_1(r,\ap)\Ep-K_2(r,\ap)E\right]\twopi, \label{eq:edp}
\end{align}
where $\Sigma$ is the surface density, $\Omega$ the orbital angular frequency, and the $K$ coefficients are related to the usual Laplace coefficients of celestial mechanics, and defined in Eq. (\ref{eq:K}). The first equation describes how the complex eccentricity of one ring evolves under the perturbation from the planet; the pressure, viscosity and self-gravity terms will be added to the right-hand side in later sections. The second equation describes how \textit{all} the rings collectively cause the eccentricity of the planet to evolve. Unless otherwise mentioned, all the integrals in this paper are carried out over the radial extent of the disc. 

We now adopt the following ansatz for the complex disc eccentricity $E(r,t)$: 
\begin{equation}
\label{eq:ansatz}
E(r,t)=f(r)~\Ed(t)=f(r)\ e_{\rm d}(t) \me^{\im\varpi_{\rm d}(t)},
\end{equation}
where $f$ is a function of radius that describes the distribution of eccentricity in the disc; it can be considered as a ``trial function'' for the disk eccentricity profile.
This ansatz {\it assumes} that the disc precesses coherently, with a common "mean disc eccentricity" $E_d(t)$ for different rings. In general, this assumption needs to be justified  a posteriori for different situations (see Section \ref{sec:discussion} for a discussion). 

We now derive the time evolution equation for $\Ed(t)$, assuming a given radial distribution $f(r)$. To achieve this, we multiply equation (\ref{eq:epd}) by $f(r)$ and integrate over the disc, yielding:
\begin{align}
\label{eq:edfr}
\dd{}{t}\int\Sigma r^2\Omega f^2(r) \Ed \twopi &= \int \im G \Mp \Sigma \big[K_1(r,\ap) f^2(r)\Ed   \nonumber \\
& -K_2(r,\ap)f(r)\Ep\big]\twopi.
\end{align}
The reason why we multiplied by $f(r)$ is related to conservation of angular momentum deficit and will become clear in Section \ref{sec:amd}.
Let us define 
\begin{align}
\Jp&=\Mp \ap^2\Op, \\
\Jd&=\int\Sigma r^2\Omega f^2(r)\twopi,
\end{align}
where $\Jp$ is the orbital angular momentum of the planet, and $\Jd$ is related to the angular momentum deficit of the disc (see Section \ref{sec:amd}). Then the equations of motion of the planet and mean disc eccentricity take the simple form:
\begin{align}
\dd{\Ed}{t}&=\im\od\Ed - \im\vd\Ep, \label{eq:edav}\\
\dd{\Ep}{t}&=\im\op\Ep - \im\vp\Ed, \label{eq:epav}
\end{align}
where
\begin{align}
\od&=\frac{1}{\Jd}\int G\Mp\Sigma K_1(r,\ap)f^2(r)\twopi, \label{omegadplanet}\\
\vd&=\frac{1}{\Jd}\int G\Mp\Sigma K_2(r,\ap)f(r) \twopi ,\\
\op&=\frac{1}{\Jp}\int G\Mp\Sigma K_1(r,\ap)\twopi ,\\
\vp&=\frac{1}{\Jp}\int G\Mp\Sigma K_2(r,\ap)f(r)\twopi .
\end{align}
We have therefore reduced the integro-differential system of Equations (\ref{eq:epd}) and (\ref{eq:edp}) to a set of two linear ODEs, one for the planet and one for the disc.
Including internal effects in the disc (pressure, self-gravity and viscosity, see Section \ref{sec:eq_disc}) will modify the $\od$ coefficient; the three other frequencies are only due to the secular planet-disc interactions, and are therefore unmodified.

\subsection{Angular momentum deficit}
\label{sec:amd}

By analogy with celestial mechanics, it is useful to consider the angular momentum deficit (AMD) of the system. For small eccentricities, the total AMD of the disc is:
\begin{equation}
\label{eq:discamd}
A_\mathrm{d} = \int\frac{1}{2}\ee\Sigma r^2\Omega\,2\pi r \,\id r.
\end{equation}
%In this Section and in Section \ref{sec:int}, all integrals are carried from $\rin$ to $\rout$, the radial extent of the disc. %With appropriate boundary conditions, the AMD is conserved for adiabatic discs.
A planet with a complex eccentricity $\Ep$ has an associated AMD of the form (for small $|\Ep|$)
\begin{equation}
\label{eq:planetamd}
A_{\rm p} =\frac{1}{2}|\Ep|^2\Mp\ap^2\Op.
\end{equation}
Secular interactions between the disc and the planet exchange AMD between the disc and the planet. The total AMD of the system, 
\begin{equation}
\label{eq:totalamd}
A = A_{\rm d} + A_{\rm p}.
\end{equation}
is a conserved quantity, that is, $\id A/\id t=0$. Multiplying equations (\ref{eq:epd}) and (\ref{eq:edp}) by the complex conjugate eccentricities and integrating eq. (\ref{eq:epd}) over the disc radial extent confirm that AMD is indeed conserved \citep{to16}.

With the disc eccentricity ansatz (\ref{eq:ansatz}), the disc-planet evolution equations (\ref{eq:edav})-(\ref{eq:epav}) satisfy 
\begin{equation}
\frac{1}{2}~\Jd~\dd{|\Ed|^2}{t}~ +~ \frac{1}{2}~\Jp~\dd{|\Ep|^2}{t}=0,
\end{equation}
which is precisely the conservation of AMD. If we had not multiply equation (\ref{eq:edfr}) by $f(r)$, the system of equations we would have eventually obtained would not have satisfied the conservation of AMD.

\section{Internal processes in the disc}
\label{sec:eq_disc}

In addition to its secular interaction with a planet, a disc will evolve under a variety of internal processes that can affect its eccentricity. Two of these processes, pressure and self-gravity, lead to precession of the disc and maintain the overall conservation of AMD. On the other hand, dissipative processes like viscosity lead to an overall damping of AMD. Resonances (Lindblad and corotation) can lead a net growth or damping of AMD, but are not included here. 

The expressions for the precession and growth rates of eccentric modes for all these processes are given in \citet{to16} and we review them here.
These integral expressions were derived assuming that the eccentricity of the system can be represented as a series of normal modes. All of these modes have a radial distribution of eccentricity given by their eigenfunction, and the introduction of the function $f(r)$ in Section \ref{sec:eq_pd} can be regarded as an ansatz of this eigenfunction. These normal modes are assumed to have a time dependence of the form $\me^{\im\omega t}$, where $\omega$ is a complex eigenfrequency. 
Such a mode corresponds to a fixed distribution of elliptical orbits of the disc and planet(s), which precess at a rate given by the real part of $\omega$ and grows (or decays) at a rate given by minus (or plus) the imaginary part of $\omega$. 

\subsection{Effect of pressure}
Pressure forces cause the disc to precess in a prograde or retrograde way. We give here the expressions for isothermal discs \citep[equivalent expressions exist for adiabatic discs, see][]{to16}, where the height-integrated 	pressure is $P=\Sigma \cs$, with $c_{\rm s}=H\Omega$ the sound speed for a disc of vertical scale-height $H$. We assume that the aspect ratio $h\equiv H/r$ is constant, and that the rotation profile is Keplerian: $\Omega=\sqrt{G\Ms/r^3}$. Using the integral relations derived in \citet{to16}, the disc precession rate due to pressure can be written in the form:
\begin{equation}
\omega_{\rm d, pr} = \frac{I_{\rm p 1} + I_{\rm p 2} + I_{\rm na} + I_{\rm 3D}}{2A},
\end{equation}
where $A$ is given by eq. (\ref{eq:totalamd}). The different terms are defined as follows:
\begin{equation}
\label{eq:intp1}
I_{\rm p1} = -\int \frac{1}{2} \Sigma \cs r^2 \left|\pd{E}{r}\right|^2 2\pi r \,\id r,
\end{equation}
\begin{equation}
\label{eq:intp2}
I_{\rm p2} = \int \frac{1}{2} \dd{(\Sigma \cs)}{r} r \left|E\right|^2 2\pi r \,\id r,
\end{equation}
\begin{equation}
\label{eq:intna}
I_{\rm na} = \int \frac{1}{2}\Sigma\dd{\cs}{r}r^2 e\pd{e}{r}\,2\pi \,\id r,
\end{equation}
(corresponding to a non-adiabatic contribution for the locally isothermal model only), and
\begin{align}
\label{eq:int3d}
I_{\rm 3D} =\int \frac{3}{2r}\Sigma\dd{}{r}(\cs r^2) \left|E\right|^2 2\pi r \,\id r,
\end{align}
which is a term arising from 3D effects described in \citet{to16}. In the 2D case we see that the two terms (Eqs.\ref{eq:intp1}--\ref{eq:intp2}) associated with pressure cause a retrograde precession of the mode, provided that the pressure gradient is negative. The 3D term (Eq. \ref{eq:int3d}) induces a prograde precession of the disc.

In \citet{to16} we also found that non-adiabatic effects damp the eccentricity of isothermal discs at a rate which depends on the gradient of the argument of pericentre, $\partial \varpi/\partial r$. Since our ansatz (Eq. \ref{eq:ansatz}) assumes rigidly precessing modes ($\partial \varpi/\partial r=0$), we do not take this effect into account. 

\subsection{Effect of self-gravity}
Self-gravity causes the disc to precess at the rate
\begin{equation}
\omega_{\rm d, sg}=\frac{I_{\rm d,sg}}{2A}
\end{equation}
with
\begin{align}
\label{eq:intdd}
I_{\rm d,sg} & = \iint \frac{1}{4}G\Sigma(r)\Sigma (r') \bigg\{ \left[ K_1 (r,r')+K_2 (r,r') \right]|E(r)-E(r')|^2\nonumber\\
& + \left[ K_1 (r,r')-K_2 (r,r') \right]|E(r)+E(r')|^2\bigg\} 2\pi r \,\id r \, 2\pi r' \,\id r'.
\end{align}

\subsection{Effect of viscosity}
Viscosity acts to damp the eccentricity. This can be written as an imaginary frequency $\im \omega_{\rm d,visc}$ with
\begin{equation}
\omega_{\rm d,visc}=\frac{J_{\rm d,visc}}{2A}
\end{equation}
and
\begin{equation}
J_{\rm visc}=\int\frac{1}{2}\ab\Sigma\cs r^2\left|\pd{E}{r}\right|^2\twopi.
\end{equation}
As discussed in \citet{to16}, this is a bulk viscosity aiming at incorporating any physical process that damps eccentricity in the disc (with the exception of resonant effects which require a specific treatment).

\subsection{Summary of all effects}

Adding all the ``internal'' disc effects to Eqs. (\ref{eq:edav})--(\ref{eq:epav}), the final set of equations for the eccentric disc-planet system becomes:
\begin{align}
\dd{\Ed}{t}&=\im(\omega_{\rm d}+ \im \omega_{\rm d,visc})\Ed - \im\vd\Ep, \label{eq:edav2}\\
\dd{\Ep}{t}&=\im\op\Ep - \im\vp\Ed, \label{eq:epav2}
\end{align}
where we have introduced 
\begin{equation}
\omega_{\rm d}=\od + \omega_{\rm d,pr} + \omega_{\rm d,sg},
\end{equation}
the free precession rate of the disc. We have therefore reduced the interaction between a planet and an extended disc to the interaction between two rings, one representing the planet and one representing the disc. The disc ``ring'' is allowed to precess on its own due to two internal forces (pressure and self-gravity), and its eccentricity can be damped due to viscosity. The system formed by Equations (\ref{eq:edav2})--(\ref{eq:epav2}) obeys the following conservation law:
\begin{equation}
\frac{1}{2}~\Jd~\dd{|\Ed|^2}{t}~ +~ \frac{1}{2}~\Jp~\dd{|\Ep|^2}{t}=-~ \omega_{\rm d,visc}~\Jd~ |\Ed|^2,
\end{equation}
where the left-hand side of this equation is the time-derivative of the (linearized) total AMD. Hence pressure and self-gravity do not affect the conservation of the total AMD, but viscosity decreases it. 
Note that mutual interactions and viscous dissipation also conserve the angular momentum of the system. This implies that the disc size will evolve. This effect is of order $e_{\rm d}^2$ and will be neglected in this paper\footnote{This situation is analogous to tidal dissipation in a multiplanet system. Consider a two-planet system (with semi-major axis $a_1<a_2$) where tidal dissipation occurs in the inner planet. The conservation of angular momentum implies that $\dot a_1/a_1=-\left[2e_1^2/(1-e_1^2)\right](\dot e_1/e_1)_{\rm tides}$ where$(\dot e_1/e_1)_{\rm tides}$ is the eccentricity damping rate due to tidal dissipation.}.

\section{Precession and damping rates for simple disc profiles}
\label{sec:toy}

We consider a disc extending from $\rin$ to $\rout$ with surface density, rotation rate and sound speed profiles given by:
\begin{align}
\Sigma(r)&=\Sigma_0~ s(r), \\
\Omega(r) &= \Oin~ \left( \frac{r}{\rin}\right)^{-3/2}, \\
c_{\rm s}(r)&=h~r\Omega,
\end{align}
where $\Sigma_0$, $\Oin$, $h (<1)$ are constant. The eccentricity profile is given by
\begin{equation}
e(r,t)=e_{\rm d}(t) f(r).
\end{equation}
With these radial profiles, the precession frequencies associated with pressure and self-gravity, and the damping rate associated with viscosity can be written as
\begin{align}
\omega_{\rm d,pr}&= h^2~\Oin ~\eta~\hat{\omega}_{\rm d,pr} \label{eq:wpr_toy}, \\
\omega_{\rm d,sg}&=\frac{\Sigma_0 \rin^2}{\Ms}~\Oin~\eta~\hat{\omega}_{\rm d,sg},\label{eq:wsg_toy},	\\
\omega_{\rm d,visc}&= \ab~ h^2~\Oin~\eta~ \hat{\omega}_{\rm d,visc}, \label{eq:wvis_toy}
\end{align}
where $\eta=(1+A_{\rm p}/A_{\rm d})^{-1}$, and the corresponding dimensionless frequencies $\hat{\omega}_{\rm d,pr}$, $\hat{\omega}_{\rm d,sg}$ and $\hat{\omega}_{\rm d,visc}$ are given in Appendix \ref{app:hatw}. The ratio of planet and disc AMDs, $A_{\rm p}/A_{\rm d}$, can be simply evaluated using Equations (\ref{eq:discamd}) and (\ref{eq:planetamd}). 

\begin{figure*}
    \begin{center}
    \includegraphics[scale=0.85,trim=0 2cm 0 0, clip]{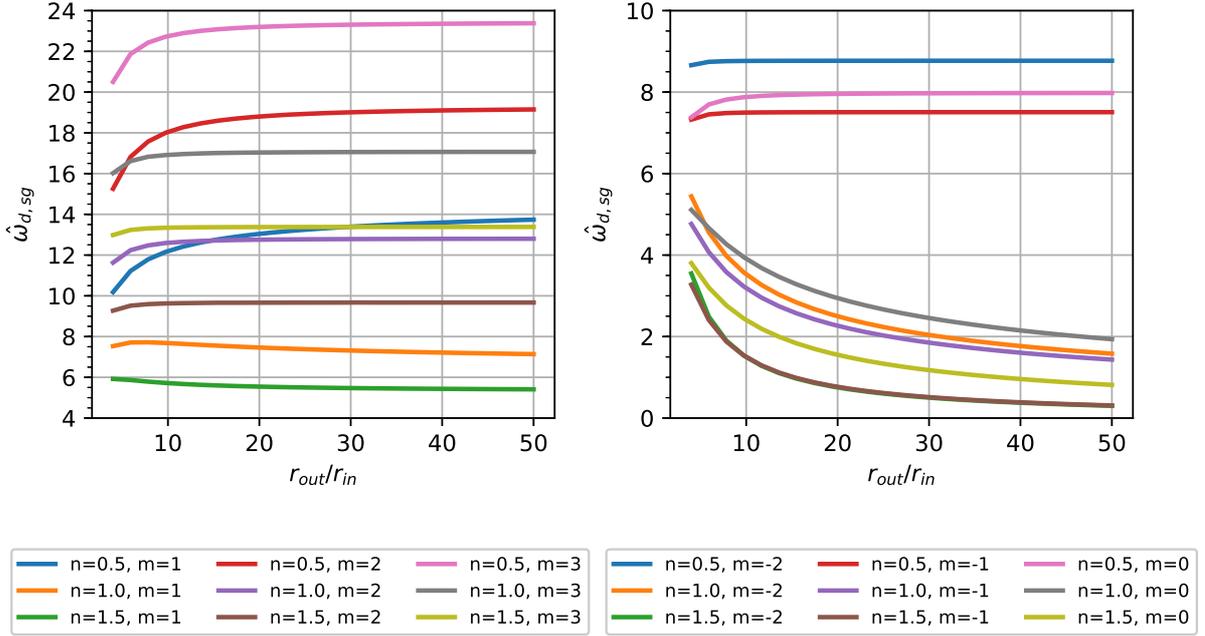}
    \caption{Value of the  self-gravity dimensionless frequency $\hat{\omega}_{\rm d,sg}$ as defined by Eq. (\ref{eq:wsg_toy}). We computed  $\hat{\omega}_{\rm d,sg}$ as a function of the disc radial extent, for different exponents $n$ and $m$ of the surface density and eccentricity power-law profiles (see Eq. \ref{eq:PLdisc}). The left panel shows the case of $m>0$ (corresponding to a profile with eccentricity larger in the inner disc), and the right panel shows the case of $m\le 0$ (corresponding to a profile with eccentricity larger in the outer disc).}
    \label{fig:sg_coeff}
    \end{center}
\end{figure*}

The dimensionless frequencies $\hat{\omega}_{\rm d,pr}$, $\hat{\omega}_{\rm d,sg}$and $\hat{\omega}_{\rm d,visc}$ depend on $\rout/\rin$, $s(x)$ and $f(x)$ (where $x\equiv x/\rin$). While $\hat{\omega}_{\rm d,pr}$ and $\hat{\omega}_{\rm d,visc}$ involve computing a simple integral, $\hat{\omega}_{\rm d,sg}$ involves a triple integral which is non-trivial to evaluate. For the convenience of readers interested in obtaining a quick ``read-out'' of $\hat{\omega}_{\rm d,sg}$, we plot it in Figure \ref{fig:sg_coeff} as a function of $\rout/\rin$ for power-law discs. We set $s(r)=(r/\rin)^{-n}$ and $f(r)=(r/\rin)^{-m}$, and compute $\hat{\omega}_{\rm d,sg}$ for different indices $(m,n)$. Note that we consider both $m>0$ and $m\le 0$. The former corresponds to a disc which is eccentric in its inner parts, as could arise when a perturber is located inside the inner edge of the disc; the latter corresponds to a disc which is eccentric everywhere or in its outer parts, as could be expected for discs perturbed by outer companions.

The precession frequencies associated with planet-disc interactions can be similarly evaluated. For concreteness we consider a planet interior to the disc ($\ap<\rin$). For $\ap\ll r$, we have $K_1\approx(3/4)\ap^2/r^3$ and  $K_2\approx(15/16)\ap^3/r^4$, and the frequencies related to disc-planet interactions can be written as
\begin{align}
\od &= \frac{3}{4} \frac{\Mp}{\Ms}\left(\frac{\ap}{\rin}\right)^2 \Oin ~\hat{\omega}_{\rm d,p} , \label{eq:wd_toy}\\
\vd &= \frac{15}{16}\frac{\Mp}{\Ms}\left(\frac{\ap}{\rin}\right)^3 \Oin ~\hat{\nu}_{\rm d,p}, \label{eq:vd_toy}\\
\op &= \frac{3}{4} \frac{M_{\rm loc}}{\Ms}\left(\frac{\ap}{\rin}\right)^3 \Op ~ \hat{\omega}_{\rm p,d},\label{eq:wp_toy} \\
\vp &= \frac{15}{16} \frac{M_{\rm loc}}{\Ms}\left(\frac{\ap}{\rin}\right)^4 \Op ~\hat{\nu}_{\rm p,d},\label{eq:vp_toy}
\end{align}
where $\Op=(G\Ms/\ap^3)^{1/2}$ and we have defined the local mass of the disc at the inner edge, $M_{\rm loc}=2\pi \Sigma_0 \rin^2$. The dimensionless frequencies $\hat{\omega}_{\rm d,p}$, $\hat{\omega}_{\rm p,d}$, $\hat{\nu}_{\rm d,p}$ and $\hat{\nu}_{\rm p,d}$ are given in Appendix \ref{app:hatw}.
Except for the dimensionless coefficients $\hat\omega$'s, these  expressions  (\ref{eq:wd_toy})--(\ref{eq:vp_toy}) are similar to those governing the secular dynamics of two planets \citep[see, e.g, Chapter 7 of][]{md99}.

\subsection{Power-law discs and inner edge effects}
\label{sec:toy_example}

\begin{figure*}
    \begin{center}
    \includegraphics[scale=0.85]{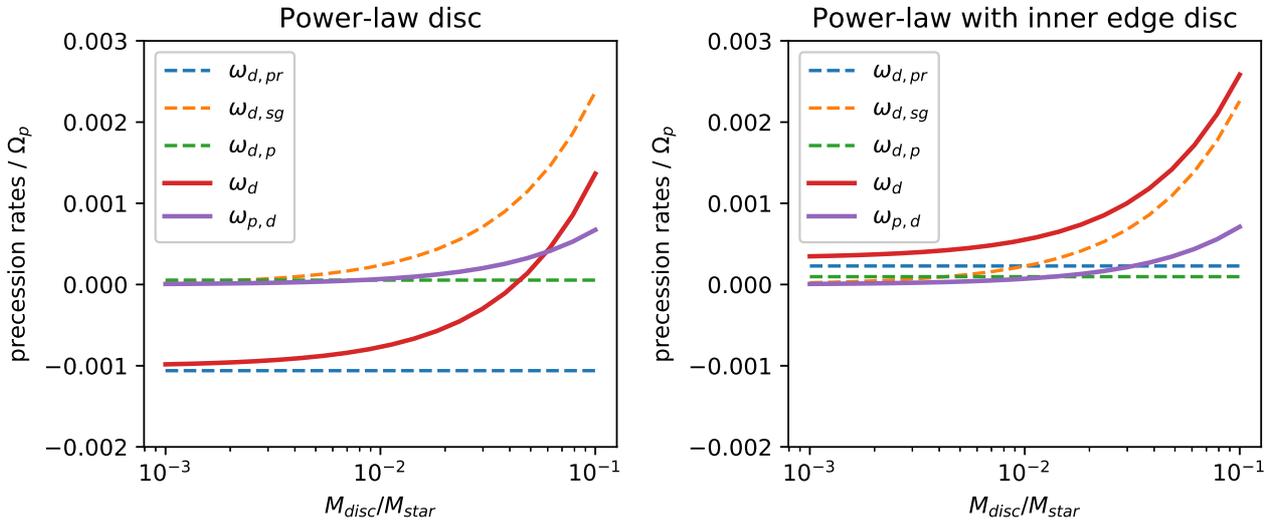}
    \caption{Various precession frequencies, in units of the planet's orbital frequency. The relevant parameters are given at the beginning of Section \ref{sec:toy_example}. The red is the total precession frequency of the disc, computed as a sum of pressure (blue), self-gravity (orange) and planet-disc interactions (green). The purple is the precession frequency of the planet induced by the disc. The \textit{left} panel shows frequencies computed with a simple power-law disc (see Eq. \ref{eq:PLdisc}). A secular resonance is possible when the red and purple curves approach each others at $M_{\rm disc}=0.05\Ms$. The \textit{right} panel shows frequencies computed with a tapering function (Equation \ref{eq:sigma_edge} with $w=0.2\ap$) that brings the surface density to 0 at the inner edge (eq. \ref{eq:sigma_edge}). The strong positive gradient at the inner edge reverses the sign of the pressure-induced precession and prevents the crossing of a secular resonance.}
    \label{fig:freqs}
    \end{center}
\end{figure*}

The simplest disc model has power-law profiles
\begin{equation}
\label{eq:PLdisc}
s(r)=\left(\frac{r}{\rin}\right)^{-n}, \qquad f(r)=\left(\frac{r}{\rin}\right)^{-m}.
\end{equation}
As an example, we consider a $1~\Mj$ planet orbiting a $1~\Msun$ star at $\ap=1~\text{au}$ on a circular orbit, and a disc extends from $\rin=2~\text{au}$ to $\rout=20~\text{au}$. We take $n=1$ and $m=3$. We impose a total disc mass and derive $\Sigma_0$ accordingly. We also assume $h=0.04$. With these profiles, we can now compute the precession frequencies $\omega_{\rm d,pr}$, $\omega_{\rm d,sg}$, $\od$ and $\op$, given by equations (\ref{eq:wpr_toy}), (\ref{eq:wsg_toy}), (\ref{eq:wd_toy}) and (\ref{eq:wp_toy}), respectively. The disc therefore precesses at a rate given by $\omega_{\rm d}=\omega_{\rm d,pr}+\omega_{\rm d,sg}+\od$. We plot these frequencies as a function of disc mass on the left panel  of Figure \ref{fig:freqs}. As the disc mass decreases (for example because of photo-evaporation), the frequencies change in time, and it is possible that the disc precession frequency matches the planet precession frequency, i.e. $\omega_{\rm d}\sim\op$. With the parameters that we chose, this happens when the disc mass is $\sim 0.05\Msun$. This suggests that as the disc dissipates, the system could cross a secular resonance.

However note that the frequency $\omega_{\rm d,pr}$ depends sensitively on the disc radial profile through the pressure gradient (Equation \ref{eq:intp2}). A realistic disc does not go to zero density suddenly at $r=\rin$. For example, when the inner planet opens a gap, the disc outside the orbit of the planet develops a strong positive density gradients. We illustrate this effect by considering the density profile:
\begin{equation}
\label{eq:sigma_edge}
\Sigma(r)= \Sigma_0 \left(\frac{r}{\rin}\right)^{-n}\frac{1}{2}\left(1+\tanh\left[\frac{\left(r-\rin\right)}{w}\right]\right),
\end{equation}
where the $\tanh$ serves as a taper function that smoothly decreases the surface density to zero around $\rin$. The steepness of the taper is specified by the width $w$.

On the right panel of Figure \ref{fig:freqs}, we re-compute the various precession frequencies, similar to the left panel of Figure \ref{fig:freqs} but with our new density profile (Eq. \ref{eq:sigma_edge}, with $w=0.1\ap$). We see that the pressure-induced precession frequency $\omega_{\rm d,pr}$ (dashed blue curve ) is significantly modified: whereas it was negative when using a power-law density profile, it is now positive. The planet's precession frequency no longer crosses the disc's precession frequency as the disc mass decreases, keeping the system away from secular resonances. 

\begin{figure}
    \begin{center}
    \includegraphics[scale=0.85]{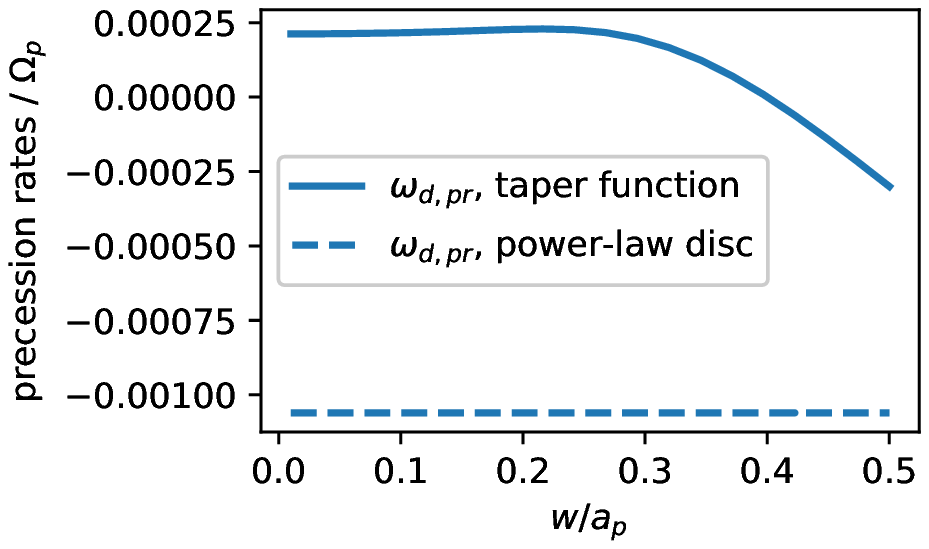}
    \caption{Variation of the pressure-induced precession rate $\omega_{\rm d,pr}$ as a function of $w$, the width of the surface density taper function (see Eq. \ref{eq:sigma_edge}). The other parameters are the same as in Fig.~\ref{fig:freqs}. For comparison, the dashed line shows the value of $\omega_{\rm d,pr}$ for a simple power-law disc (see Eq. \ref{eq:PLdisc}).}
    \label{fig:freqs_w}
    \end{center}
\end{figure}

On Figure \ref{fig:freqs_w}, we plot $\omega_{\rm d,pr}$ as a function of $w$, the width of the taper function, with all the other parameters kept the same as above. For $w\lesssim0.25\ap$ (corresponding to steep density gradients), the value of $\omega_{\rm d,pr}$ is mostly unaffected by the choice of $w$. As $w$ increases (and the gradient decreases), $\omega_{\rm d,pr}$ changes sign from positive to negative.

\section{One planet and an outer disc}
\label{sec:ragusa}

In this section we apply our simplified eccentric disc dynamics model to study the  long-term evolution of a giant planet interacting with an outer disc. We compare our result to the hydrodynamical simulations of \citet{ragusa18} and show that our simplified model can reproduce the main secular features of the simulation results.

\subsection{Main results of the \citet{ragusa18} simulations}
\label{sec:ragusa_sim}

\citet{ragusa18} carried out two long-term simulations (about 300,000 planetary orbits) of a giant planet ($\Mp= 13\Mj$) in a disc. 
In their simulations, the disc was set up to be locally isothermal, with a radially-dependent aspect ratio: $h=H/r=0.036(r/\ap)^{0.215}$, extending from $\rin=0.2\ap$ to $\rout=15\ap$. The initial surface density was set to a power-law profile $(r/\ap)^{-0.3}$ with an exponential taper for $r>5\ap$. The viscosity was implemented with an $\alpha$-prescription, where $\alpha=10^{-3}(r/\ap)^{-0.63}$ also varies with radius. The boundary was closed at the outer edge of the disc, and open at the inner edge. The planet and the disc were initially on circular orbits, and the planet was free to evolve under the combined gravitational potential of the star and disc.  \citet{ragusa18} considered two cases: one with a  ``light'' disc of initial mass $0.2 \Mp$, and another with a ``massive'' disc of mass $0.65 \Mp$.
In the simulations, the inner disc drains on a relatively short timescale, leaving behind a planet and outer disc. For the ``light'' disc case, the system eventually settles in an aligned mode, where the disc rigidly precesses at the same rate as the planet, so that the difference in arguments of pericentres between the planet and the disc librates around 0. The eccentricity of the disc grows rapidly to reach 0.1 at the location where the radial profile of disc AMD is maximum. The eccentricity of the planet grows much more slowly, to reach 0.12 after 300,000 orbits. For the ``massive'' disc case, the system settles in an anti-aligned mode, where the arguments of pericentre of the rigidly-precessing disc is 180\textdegree away from that of the planet. The disc and the planet initially experience a rapid growth of eccentricity (up to $\sim 0.1$), before slowly decaying over  time. 

Although the rapid growth of eccentricity at the beginning of each simulations \citep[which is due to Lindblad resonances and has been studied extensively by][]{to17} cannot be captured by our model, the secular evolution and (possible) viscous damping seen in the simulations can serve as a benchmark to test the validity of our model. In particular, we focus on the locking of the system in either the aligned or anti-aligned mode. 

\subsection{Result from eccentric disc model and comparison with simulations}
\label{sec:ragusa_disc}
Since the hydrodynamical simulations of \citet{ragusa18} were 2D, we set the 3D term in Equation (\ref{eq:edav2}) to be 0. We also ignore the effect of self-gravity. As the disc interior to the planet's orbit was rapidly cleared away in the simulations, we focus on the effect of the outer disc.

\begin{figure}
    \begin{center}
    \includegraphics[scale=0.65]{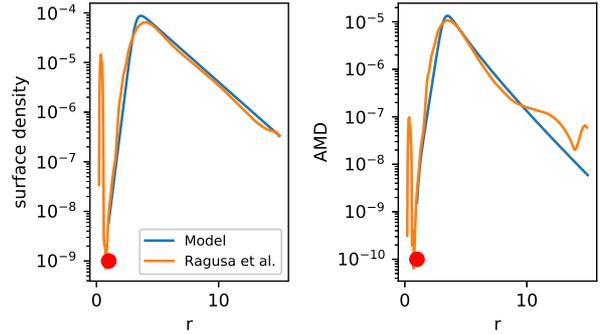}
    \caption{Surface density (left) and AMD (right) profiles for the disc model of \citet{ragusa18}. The orange curves are directly extracted from the simulations after the system has experienced eccentricity growth due to resonances, while the blue curves represent our disc model. The disc AMD profile is given by the integrand of Eq. (\ref{eq:discamd}). Our disc model does not attempt to include the disc inside the orbit of the planet, whose position is indicated by a red dot.}
    \label{fig:ragusa_disc}
    \end{center}
\end{figure}

We find that using the following surface density profile:
\begin{equation}
\label{eq:sigma_rag}
\Sigma(r)= \Sigma_0 \me^{-(1.6r/\rin)}~\frac{1}{2}\left(1+\tanh\left[\frac{\left(r-\rin\right)}{w}\right]\right),
\end{equation}
with $\rin=3.\ap$ and $w=0.3\ap$ matches well the simulation's density profile for the disc exterior to the planet at $t=30,000$ orbits, as shown in Figure \ref{fig:ragusa_disc}.
At this point of the simulation, the disc's eccentricity has grown and saturated to its maximum value, and the disc-plane system in interacting secularly; it its therefore a good starting point for our model.
For the eccentricity function $f$, we use the simple power law $f(r)=r^{-3/2}$. Combined with the surface density profile, this eccentricity profile gives a good match for the radial distribution of disc AMD, as can be seen in the right panel of Figure \ref{fig:ragusa_disc}. 
We also set $\alpha=0.025$ and $H/r=0.036$. The initial complex eccentricities of the disc and planet match that of the simulations at 30,000 orbits, and are given in Table \ref{tab:ragusaIC}.

\begin{table}
\centering
\caption{Initial conditions for the simulations presented in Section \ref{sec:ragusa_disc}}.
\label{tab:ragusaIC}
\begin{tabular}{lcc}
\hline
 & Light disc & Massive disc\\
 \hline
 $\ep$ &  0.05 & 0.1 \\
 $e_{\rm d}$ & 0.1 & 0.14 \\
 $\varpi_{\rm p} $ & 0\textdegree & 0\textdegree \\
 $\varpi_{\rm d} $ & 50\textdegree & 160\textdegree \\
\hline
\end{tabular}
\end{table}

\begin{figure*}
    \begin{center}
    \includegraphics[scale=0.7]{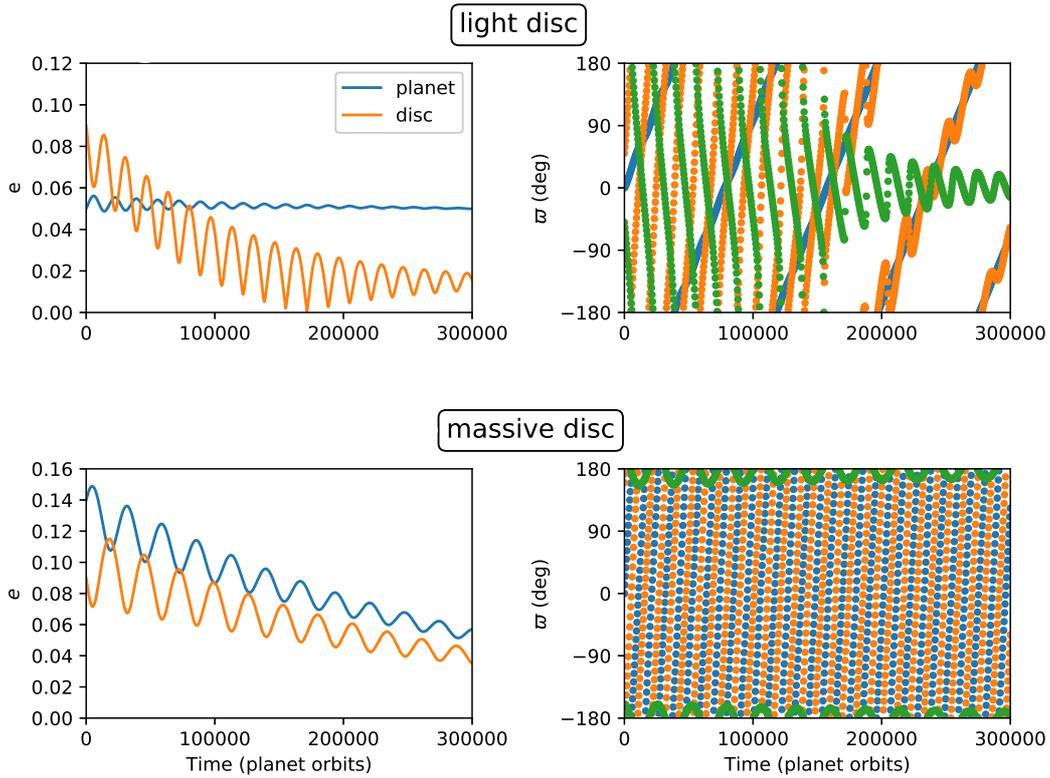}
    \caption{Time evolution of the planet (blue) and disc (orange) eccentricity (left) and argument of pericentre (right), for the ``light disc'' (top ) and ``massive disc'' (bottom) case of \citet{ragusa18}, but calculated using our simplified eccentric disc model.. The disc's eccentricity and argument of pericentre are measured at the location where the disc AMD profile is maximum (see the right panel of Figure \ref{fig:ragusa_disc}). On the right panels, the green points give the difference between the arguments of pericentre of the planet and the disc. The parameters of these calculations are described in Section \ref{sec:ragusa_disc}.
    }
    \label{fig:ragusa_eigv}
    \end{center}
\end{figure*}

The top panels of Figure \ref{fig:ragusa_eigv} show the result of our calculation when applied to the ``light disc case'', while the bottom panels shows the result for the ``massive disc'' case. These results are to be compared with the left and right panels of Figures 6 and 7 of \citet{ragusa18}, respectively. The qualitative agreement between our model and and hydrodynamical simulations is good, and we proceed to quantify this further in the next section.

\subsection{Normal mode interpretation}
\label{sec:modes}
Equations (\ref{eq:edav2}) and (\ref{eq:epav2}) can be solved as a superposition of two eigenvectors, each of them having an associated eigenfrequency. One mode is ``aligned'' (meaning that the difference of arguments of pericentres of the planet and disc is zero), while the other one is ``anti-aligned'' (the difference is 180\textdegree). In addition, the aligned mode precesses slower than the anti-aligned. More details, as well as a discussion of adding the effect of damping on these modes, can be found in \citet{zhang13}. Let's denote the aligned and anti-aligned eigenvectors by $\mathbf{v}_+$ and $\mathbf{v}_-$, respectively, with associated eigenfrequencies $g_+$ and $g_-$. Because of viscosity, the eigenfrequencies have a real part (associated with precession) and an imaginary part (associated with viscous damping). Then the time evolution of the disc and planet complex eccentricities is given by:
\begin{align}
\begin{pmatrix}
\Ed \\
\Ep
\end{pmatrix}
=A \mathbf{v}_+
\me^{\im g_+ t}
+
B\mathbf{v}_-
\me^{\im g_- t}
\end{align}
where $A$ and $B$ are constant determined by initial conditions. In Table \ref{tab:eigv} we give the values of the eigenvectors and eigenfrequencies for both the light and massive disc cases. The light disc shows a clear distinction between the aligned and anti-aligned mode, with the aligned mode precessing 8 times slower, and being damped 50 times slower than the anti-aligned mode. As a consequence, the light disc case presented in Figure \ref{fig:ragusa_eigv} naturally ends up in the aligned configuration. The anti-aligned mode is predominantly in the disc, which explains why the eccentricity of the disc decreases over the timescale of the light disc simulation, while that of the planet remains roughly constant. We have verified that regardless of the initial differences in the arguments of pericentres (ranging from 0\textdegree~ to 180\textdegree), the system always ends up in the aligned configuration within $3\times10^5$ orbits for our choice of fiducial viscosity. 

In the massive disc case, the modal structure is rather different. Both modes are roughly equally shared between the planet and the disc, and have  similar precession and damping rates. The anti-aligned mode damps slightly slower, and we expect that the system would settle in this configuration given enough time, regardless of the initial conditions. We have verified this by starting the the massive disc calculation with the disc and the planet aligned. The system slowly departs from this configuration, and after about $10^6$ orbits it is fully in the anti-aligned configuration.

The normal mode analysis gives insight into the dynamics in the presence of damping. Depending on the disc mass, the system settles in either the aligned or anti-aligned mode, and more massive discs always lead to anti-alignment. Overall, our eccentric disc model successfully captures the secular evolution of the \citet{ragusa18} simulations.

\begin{table*}
\centering
\caption{Eigenvectors and eigenvalues of a ``planet + outer disc'' system, for  the light and massive disc cases (see Section \ref{sec:ragusa}). The eigenfrequencies are given in units of the planet's orbital frequencies.}
\label{tab:eigv}
\begin{tabular}{lcccc}
\hline
 & Aligned eigenvector ($\mathbf{v}_+$) & Aligned eigenvalue ($g_+$)  & Anti-aligned eigenvector ($\mathbf{v}_-$) & Anti-aligned eigenvalue ($g_-$) \\
\hline
Light disc & 
 $\begin{pmatrix}
0.31-0.01~\im \\
0.95
\end{pmatrix}$ & $1\times 10^{-5}+2\times 10^{-8}~\im$  &
$\begin{pmatrix} 
1 \\
-0.05+10^{-3}~\im
\end{pmatrix}$
& $8\times 10^{-5}+ 10^{-6}~\im$  
 \\
Massive disc& $\begin{pmatrix}0.82 \\
0.57+0.02~\im
\end{pmatrix}$ & $6\times 10^{-5}+8\times 10^{-7}~\im$
 & $\begin{pmatrix}
-0.63-0.02~\im\\ 
0.77
\end{pmatrix}$ & $9\times 10^{-5}+5\times 10^{-7}~\im$  \rule{0pt}{6ex} \\
\hline
\end{tabular}
\end{table*}

\subsection{Discussion}
\label{sec:ragusa_discussion}
Although we obtain a qualitative match with the two simulations presented in \citet{ragusa18}, some differences exist. A quantitative match cannot be obtained as the surface density evolves in time in the hydrodynamical simulations. In addition, a shift from prograde to retrograde precession occurs in the later part of the massive disc simulation of \citet{ragusa18}, which we could not reproduce nor explain.

We had to employ an $\alpha$ viscosity ($\ab=0.025$) to reproduce the mode damping rate seen in the \citet{ragusa18} simulations. This is  higher than the viscosity used by \citet{ragusa18}. This discrepancy arises because  our model uses a \textit{bulk} viscosity. As stressed by \citet{to16}, this bulk term represents whichever process (apart from resonances) acts to damp the eccentricity, and should not be strictly regarded as an hydrodynamical viscosity.

The simulations of \citet{ragusa18} capture more physics, as they allow for the eccentricities to grow under the effect of Lindblad resonances, in agreement with the theory of \citet{to16}. In the light disc simulation of \citet{ragusa18}, this led to a slow growth of eccentricity of the planet from 0 to 0.1 over 300,000 orbits. Our secular disc model does not capture this. In most cases however, the growth of eccentricity of the disc due to Lindblad resonances is very rapid and saturates to moderate values, as studied in \citet{to17}. It is therefore acceptable to start our simulations with an eccentric disc, and assume that the planet is slightly eccentric.

\section{Two planets and an outer disc}
\label{sec:kepler419}

As an application of our simplified eccentric disc evolution model, we consider a system of two giant planets interacting with an outer disc. \citet{petrovich19} recently studied the Kepler-419 system, where two giant planets orbit a central star in an anti-aligned configuration ($\varpi_c-\varpi_b\simeq 182^\circ$) with high eccentricities \citep[$e_b\simeq 0.82$ for the inner planet, and $e_c\simeq 0.18$ for the outer planet; see][]{dawson14}. They proposed a model in which disc clearing over long time-scales (for instance due to photo-evaporation) causes the two planets to cross a secular resonance which would excite their eccentricities to large values and lock them in an anti-aligned configuration, in agreement with observations. However, \citet{petrovich19} treated the disc as being passive, with no eccentricity; in essence, they did not take into account the back reaction of the planets on the disc, and the disc serves as an infinite reservoir of angular momentum. In this section we use our simplified eccentric disc evolution model to explore whether the dynamics observed by \citet{petrovich19} still holds when the planets and the disc are allowed to exchange eccentricities.

\subsection{Initial conditions and disc model}
\label{sec:kepler419disc}
We consider the evolution of two planets and a disc, with the two planets having similar properties as the Kepler-419 system: $M_{\rm b}=2.77\Mj$ and $M_{\rm c}=7.65\Mj$, $a_{\rm b}=0.3745~\text{au}$ and $a_{\rm c}=1.697~\text{au}$, orbit a star of mass $\Ms = 1.438\Msun$. 
Following \citet{petrovich19}, we use the following initial conditions: $e_{\rm b}=0.05$, $e_{\rm c}=0.4$, and $\varpi_{\rm c}-\varpi_{\rm b}=150\degree$. Unlike \citet{petrovich19}, we do not consider the inclination evolution. In addition to planet-disc interactions, we must also include the interaction between planet b and planet c. The relevant equations governing the planet-planet interactions are given in Appendix \ref{app:pp}.

We use a similar disc model as \citet{petrovich19}. The disc surface density is given by Equation (\ref{eq:sigma_edge}) with $n=3/2$. The gap width $w=0.1a_{\rm c}$ is such that the disc radial AMD profile is maximum at $r=1.5 a_{\rm c}$, and the disc extends to $\rout=15 a_{\rm c}$. The disc mass is initially set to $M_{\rm d,0}=0.1 \Ms$, and decreases in time as
\begin{equation}
\Md(t)=M_{\rm d,0}/(1+t/\tau_{\rm v}),
\end{equation}
where $\tau_{\rm v}=10^5~\text{yr}$ is the disc decay time. We also assume $h=H/r=0.05$ and $\ab=0.01$. The disc eccentricity profile is a power-law with $m=3$ (see Eq. \ref{eq:PLdisc}) and the initial disc eccentricity is normalized so that $e_{\rm d}=0.1$ at the location in the disc where the radial profile of AMD is maximum.

\subsection{Results}
\label{sec:kepler419res}
\begin{figure*}
    \begin{center}
    \includegraphics[scale=0.7,]{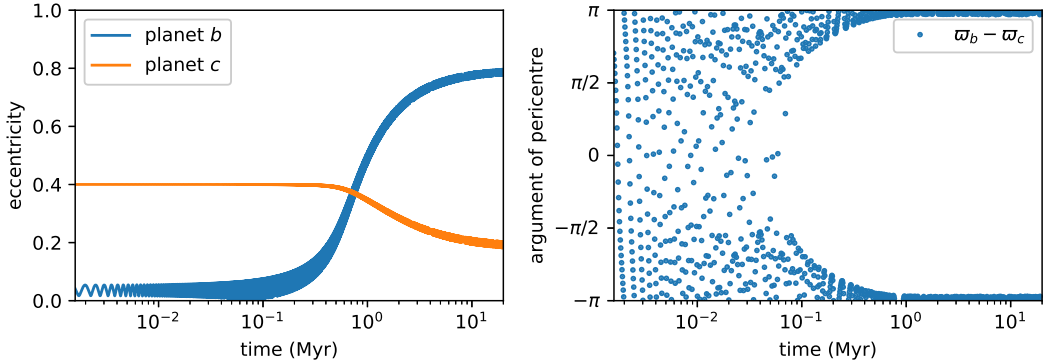}
    \caption{Time evolution of the eccentricity (left) and argument of pericentre (right) of two planets similar to that of the Kepler-419 system, under the same assumption as \citet{petrovich19}.}
    \label{fig:P19noBR}
    \end{center}
\end{figure*}

We first perform a similar calculation as \citet{petrovich19}, assuming  that the disc has no eccentricity, does not precess, nor viscously decays in time (i.e., $\omega_{\rm d,pr}=\omega_{\rm d,sg}=\omega_{\rm d,visc}=0$), and that the planet does not back-react on the disc, (i.e., $\omega_{\rm d,p}=\nu_{\rm d,p}=0$). 
In Figure \ref{fig:P19noBR} we show the result of such calculation. We see that after a few $10^5~\text{yrs}$, the system crosses a secular resonance, which excites the eccentricity of planet $b$  while decreasing the eccentricity of planet $c$. The final planetary eccentricities are near the observed values, and the arguments of pericentres are locked into an anti-aligned configuration. Our Figure \ref{fig:P19noBR} exhibits a similar evolution and end result as that shown in Figure 4 of \citet{petrovich19}, despite the fact that we do include inclination evolution.

\begin{figure*}
    \begin{center}
    \includegraphics[scale=0.7,]{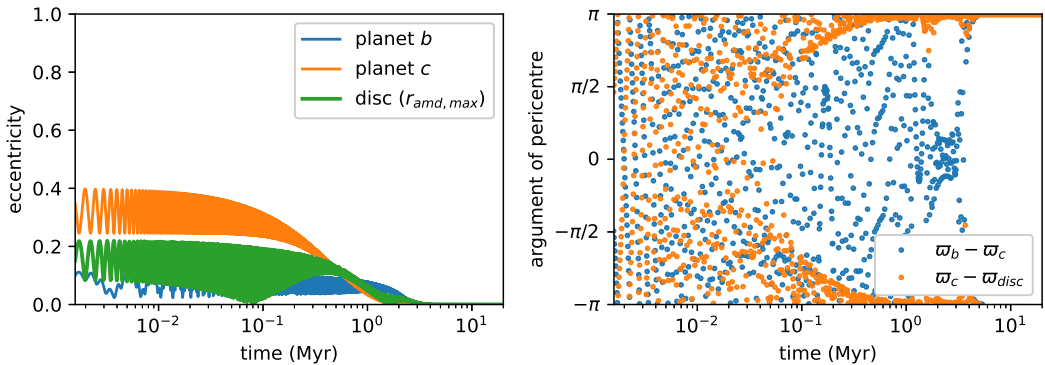}
    \caption{Similar to Figures \ref{fig:P19noBR}, but including the back reaction on the disc from the planets, as well as the pressure, viscosity, and self-gravity effects in the disc. The green curve on the left panel is the eccentricity of the disc, measured at the location of maximum AMD profile in the disc.}
    \label{fig:P19BR}
    \end{center}
\end{figure*}

Next we proceed to examine the consequences of allowing back-reaction of the planet on the disc, as well as allowing the disc eccentricity to evolve under its own internal effects. 
In Figure \ref{fig:P19BR} we show that in this case, the mechanism observed by \citet{petrovich19} no longer operates, and the eccentricities are damped. The outer planet and the disc become locked in an anti-aligned configuration, indicating that they are likely to be evolving in a common secular mode. Because of this mode coupling, damping of the disc eccentricity due to viscosity propagates to the rest of the system and eventually damps the eccentricities of the two planets. 

We have experimented with different values of $h$, as well as different eccentricity distribution profiles (different power-law profiles) and different values of the coefficients that determine the width of the surface density tapering function (see Section \ref{sec:toy_example}). In no case were we able to reach the high planetary eccentricities attained in the \citet{petrovich19} calculation.

\begin{figure*}
    \begin{center}
    \includegraphics[scale=0.7,]{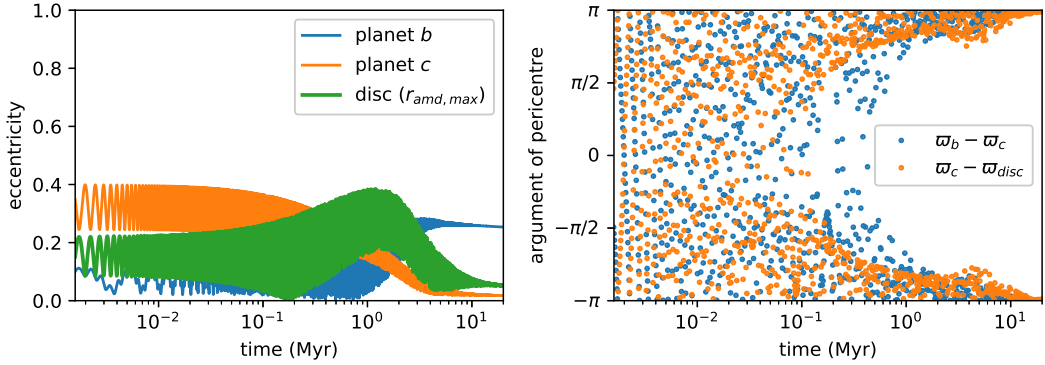}
    \caption{Similar to Figures \ref{fig:P19BR}, but with a lower viscosity of $\ab=0.001$}
    \label{fig:P19BRa001}
    \end{center}
\end{figure*}

The disc viscosity is uncertain. With our canonical choice of $\ab=0.01$ in Figure \ref{fig:P19BR}, we managed to entirely suppress the growth of planetary eccentricity. In Figure \ref{fig:P19BRa001} we show the same calculation as in Figure \ref{fig:P19BR}, but with $\ab=0.001$. In this case the eccentricity excitation of planet b, although quenched, is not fully suppressed, and the final eccentricity reaches $e_{\rm b}\sim 0.25$. In the meantime the disc goes through a phase of high eccentricity before being damped. Hydrodynamical simulations indicate that growth of eccentricity in discs tends to saturate when the maximum eccentricity reaches 0.2--0.3, at which point non-linear effects related to strong pressure gradients and nearly crossing orbits prevent further eccentricity growth \citep{to17}. These non-linear effects are not included in our model. The high disc eccentricity phase seen in Figure \ref{fig:P19BRa001} is a consequence of the crossing of a secular resonance: at the time of crossing, the disc mass is about $3\Mj$, similar to planet b, and lower than planet c; thus the disc is susceptible to strong eccentricity excitation, because it no longer strongly dominates the angular momentum budget of the system. This is one of the limitations of our model, where a low-mass disc might be subject to unphysically high excitation of eccentricity, especially if resonant crossing is involved.  In reality, once the disc reaches a high eccentricity, nonlinear dissipation will likely lead to significant damping, and the final planet b eccentricity will be lower than depicted in Figure \ref{fig:P19BRa001}.

\section{Discussion}
\label{sec:discussion}

Our simplified theory and equations for the secular dynamics of
eccentric discs developed in Sections \ref{sec:eq_pd}-\ref{sec:eq_disc} are necessarily
approximate. They assume small disc eccentricity and do not account
for the possibility of eccentricity excitation due to Lindblad
resonances. The eccentricity profile $f(r)$ (``trial function'') must be prescribed 
in the theory. Most importantly, the theory assumes that the disc as a whole
precesses coherently, with different ``rings'' having the same
precession rate (see Eq.~\ref{eq:ansatz}).

Under what conditions is the ``coherent disc'' ansatz valid? While a
rigorous answer to this question can only come from hydrodynamical
simulations or from solving the complicated integral-differential eccentric disc
equations of \citet{to16} (which also do not take account of
non-linear effects), we can offer the following qualitative answer.
Consider a disc interacting with an interior planet ($a_p<r_{\rm in}$). 
The torque (per unit mass) acting on the disc from the planet is $T_{\rm p}\sim
GM_p a_p^2/r^3$, and it gives rise to differential apsidal precession
$\omega_{\rm p} \sim T_{\rm p}/(r^2\Omega)\sim (M_p/M_\star)(a_p/r)^2\Omega$.
On the other hand, the torques due to disc pressure and self-gravity are
$T_{\rm pr}\sim c_s^2$ and $T_{\rm sg}\sim \pi G\Sigma r$, respectively, 
and they keep the disc coherent at the rates\footnote{Note that for a Keplerian disc, eccentric waves propagate at a speed of order $v_{e}\sim c_s(H/r)$, so the ``communication time'' due to pressure 
is $r/v_e\sim \omega_{\rm pr}^{-1}$.}
given by $\omega_{\rm pr}\sim (H/r)^2\Omega$ and
$\omega_{\rm sg}\sim (\pi \Sigma r^2/M_*)\Omega$. A necessary condition
for the disc to maintain coherence is that 
$\omega_{\rm pr}\gtrsim \omega_{\rm p}$ or $\omega_{\rm sg}\gtrsim \omega_{\rm p}$ 
is satisfied. For discs with $H/r\sim $constant, the condition $\omega_{\rm pr}\gtrsim 
\omega_{\rm p}$ can be more easily satisfied at large radii than at small radii.

A more stringent condition for disc coherence is that the global disc precession time $|g|^{-1}$
(where $g$ is the eigenfrequency of the disc-planet system; see Section \ref{sec:modes} for example)
be longer than the ``eccentricity communication times'' due to pressure and self-gravity 
throughout the disc. That is, coherent precession of the disc requires
\begin{equation}
\omega_{\rm pr}\sim (H/r)^2\Omega \gtrsim |g|,~~~{\rm or}~~~
\omega_{\rm sg}\sim (\pi\Sigma r/M_*)\Omega \gtrsim |g|.
\end{equation}
Since both $\omega_{\rm pr}$ and $\omega_{\rm sg}$ decrease with increasing $r$,
these conditions inevitably break down at large disc radii.

In practice, for an extended disc, a precise determination of the
coherent disc region is not crucial for capturing the secular effect
of eccentric disc - planet interaction in our simplified theory. As we
see from Sections \ref{sec:eq_pd}-\ref{sec:eq_disc}, the disc eccentricity ``trial function'' $f(r)$
enters into the expressions of all relevant disc precession
frequencies. For a sufficiently rapid declining function $f(r)$, these
frequencies are not modified significantly by the choice of the outer
(coherent) disc radius, as long as the inner disc region, where most
of the disc AMD is concentrated, is coherently precessing.
In another word, we can choose the function $f(r)$
to vanish in the non-coherent region and the result will not be
affected. In discs tidally perturbed by a planet/companion, the fastest
growing eccentric mode is expected to develop near the perturber and
to damp far away from it, so choosing a function $f$ that vanishes at
one end of the disc is physically justified \citep{lubow1991a,to16}.

Another uncertainty in applying our simplified eccentric disc
evolution equations arises from the pressure-induced precession
term. We have seen in Section \ref{sec:toy_example} that $\omega_{\rm
  d,pr}$ can depend sensitively on the disc pressure gradient, and
introducing a taper function that brings the surface density to zero
at the inner disc edge can significantly change the disc precession
rate, which in turn affects the possibility of secular resonance
crossing. The parameters that govern the shape of this taper are
somehow arbitrary, and different values will affect the result, mostly
by changing the slope of the pressure gradient at the inner edge. The
values we chose in Sections \ref{sec:ragusa} and \ref{sec:kepler419} gave a good fit to the simulations by
\citet{ragusa18}, but a better understanding of the disc truncation at
the inner edge would lead to a more robust result.

In the application discussed in Section \ref{sec:kepler419}, we have followed 
\citet{petrovich19} to consider a simple disc clearing model
--- this has allowed us to illustrate the important role
played by including the dynamics of eccentric discs.
Obviously, realistic disc clearing is a much more complicated process, involving 
accretion onto the star, photo-evaporation and magnetic winds.
One possibility that we have not addressed in this paper
is that the inner disc radius could change in time during disc clearing,
thereby further reducing the strength of planet-disc interactions.

One additional caveat concerns our treatment of planet-planet interactions
in the application discussed in Section  \ref{sec:kepler419}. Our equations for
planet-disc interactions are linear and valid for small eccentricities
only. On the other hand, we have modelled planet-planet interactions
using an expansion in semi-major axis, valid for arbitrary
eccentricities.  Strictly speaking, this non-linear expansion does not
conserve the linear AMD introduced in Section \ref{sec:amd}. We have
found that this inconsistency does not significantly affect our
results. One could use the linear Laplace-Lagrange theory to model
planet-planet interactions, which has a similar structure as our
planet-disc equations, and would conserve linear AMD exactly. However,
its accuracy when a planet reaches a large eccentricity would be
doubtful.  We have also investigated whether introducing non-linear
terms in the planet-disc equations lead to significant different
results. In order to do so, we made use of the formalism developed by
\citet{ogilvie07} to compute higher order terms in the disturbing
function. We obtained evolutionary equations with terms proportional
to $E^3$, which turned out to have a negligible effect on our results. We
chose not to include these rather lengthy equations in this paper, and
to neglect these non-linear effects.

Our treatment of self-gravity relies on a softening of the Laplace
coefficients. Softening schemes in self-gravitating discs have been
studied by many authors
\citep[e.g.,][]{tremaine01,touma02,hahn03,to16,sr19,lee19}. Recently,
\citet{sr19} examined how applying the softening directly to the disc
potential results in modified expressions for the Laplace
coefficients. An interesting feature of this approach is that it gives
rise to the possibility that the disc precesses retrogradely under its
own self-gravity, while the approach we use here \citep[based
  on][]{to16} always leads to prograde precession.\footnote{In fact, a calculation
of the linear eccentricity eigenmodes of self-gravitating discs (treated as a 
dense nested rings) shows that all modes have prograde procession, regardless of the disc
surface density profile (Bonan Pu, private communication).}
While the approach
of \citet{sr19} might be true for particle (e.g., debris) discs, it is
not clear whether it still holds true for gaseous discs, where
pressure prevents streamline crossing and provides a natural softening
length. Recent numerical \citep{mutter17} and analytical works
\citep{lee19} indeed find that discs supported by pressure and
self-gravity precess in the prograde direction.

Finally, it is of interest to compare our planet - eccentric disc
model with planet - inclined disc models studied by many authors. The
theory of warped discs in the presence of an inclined perturber is as
complex as that of eccentric discs, if not more so \citep[see, e.g.,][]{pp83,pl95,ogilvie99,lo00,ogilvie01,lo01,ogilvie06,ol13,fl14,ogilvie18}. However, if we assume
that the disc warp is small and that the disc is able to maintain
coherent rigid-body-like nodal precession --- both assumptions can be
justified in many situations, then the evolution equations governing
an inclined planet - disc system become very simple, and can lead to
many novel applications \citep[see, e.g.,][]{batygin12,lai14,zl18a,zl18b,zl18c}. The simplified planet - eccentric disc model developed in this
paper has similar virtues as the ``inclined planet - rigid disc''
model. Thus, despite the various caveats discussed above, the model
developed in this paper should find useful applications in many planet
- disc problems (and Sections \ref{sec:ragusa} and \ref{sec:kepler419} clearly illustrate the utility
of our model).

%%%%%%%%%%%%%%%%%%%%%
\section{Conclusion}
\label{sec:conclusion}

In this paper we have developed a simplified model of planet-disc
interactions that captures the long-term (secular) dynamics of the
eccentric disc and planet(s).  The model is based on the ansatz that
the disc maintains coherent apsidal precession under the combined
influences of the external (from the planet) and internal forces. Our
model includes the gravitational coupling between the disc and the
planet(s), as well as internal effects due to pressure, self-gravity
and viscosity. Equations (\ref{eq:edav2}) and (\ref{eq:epav2}) are the
main result of the paper. They reduce the planet-disc eccentricity
evolution to a set of linearized ODEs, where the planet and the disc
are each represented by an eccentric ring. This set of equations can
be generalized to multiple planets. The equations are easy to
implement, and can be used study the long-term eccentricity evolution
of planet-disc systems without resorting to computationally demanding
hydrodynamical simulations.

Applying our equations to the ``giant planet + external disc'' system,
we found that our model captures the main secular features seen in the
long-term hydrodynamical simulations of \citet{ragusa18}. We also
revisited the work of \citet{petrovich19} in light of our improved
treatment of planet-disc interactions. We found that the eccentricity
growth observed when two planets cross a secular resonance while an
external disc is dispersing can be significantly quenched when the
back reaction of the planets on the disc is allowed. This illustrates
the importance of properly taking account of the dynamics of eccentric
discs in studying ``planets + disc'' systems.

While our model and equations are approximate and do not include some 
important effects (such as eccentricity excitation and non-linear effects),
they can be easily adapted to any astrophysical problems involving 
eccentric discs and perturbing companions. 

\section*{Acknowledgements}
We thanks Bonan Pu for useful discussion. This work has been
supported in part by NASA grants NNX14AG94G and 80NSSC19K0444, and NSF
grant AST-1715246.

\bibliographystyle{mnras}

\bibliography{biblio}

\begin{thebibliography}{}
\makeatletter
\relax
\def\mn@urlcharsother{\let\do\@makeother \do\$\do\&\do\#\do\^\do\_\do\%\do\~}
\def\mn@doi{\begingroup\mn@urlcharsother \@ifnextchar [ {\mn@doi@}
  {\mn@doi@[]}}
\def\mn@doi@[#1]#2{\def\@tempa{#1}\ifx\@tempa\@empty \href
  {http://dx.doi.org/#2} {doi:#2}\else \href {http://dx.doi.org/#2} {#1}\fi
  \endgroup}
\def\mn@eprint#1#2{\mn@eprint@#1:#2::\@nil}
\def\mn@eprint@arXiv#1{\href {http://arxiv.org/abs/#1} {{\tt arXiv:#1}}}
\def\mn@eprint@dblp#1{\href {http://dblp.uni-trier.de/rec/bibtex/#1.xml}
  {dblp:#1}}
\def\mn@eprint@#1:#2:#3:#4\@nil{\def\@tempa {#1}\def\@tempb {#2}\def\@tempc
  {#3}\ifx \@tempc \@empty \let \@tempc \@tempb \let \@tempb \@tempa \fi \ifx
  \@tempb \@empty \def\@tempb {arXiv}\fi \@ifundefined
  {mn@eprint@\@tempb}{\@tempb:\@tempc}{\expandafter \expandafter \csname
  mn@eprint@\@tempb\endcsname \expandafter{\@tempc}}}

\bibitem[\protect\citeauthoryear{{Batygin}}{{Batygin}}{2012}]{batygin12}
{Batygin} K.,  2012, \mn@doi [\nat] {10.1038/nature11560}, \href
  {http://ui.adsabs.harvard.edu/abs/2012Natur.491..418B} {491, 418}

\bibitem[\protect\citeauthoryear{{Brouwer} \& {Clemence}}{{Brouwer} \&
  {Clemence}}{1961}]{BC61}
{Brouwer} D.,  {Clemence} G.~M.,  1961, {Methods of celestial mechanics}.
Academic Press

\bibitem[\protect\citeauthoryear{{Dawson} et~al.,}{{Dawson}
  et~al.}{2014}]{dawson14}
{Dawson} R.~I.,  et~al., 2014, \mn@doi [\apj] {10.1088/0004-637X/791/2/89},
  \href {http://ui.adsabs.harvard.edu/abs/2014ApJ...791...89D} {791, 89}

\bibitem[\protect\citeauthoryear{{Foucart} \& {Lai}}{{Foucart} \&
  {Lai}}{2014}]{fl14}
{Foucart} F.,  {Lai} D.,  2014, \mn@doi [\mnras] {10.1093/mnras/stu1869}, \href
  {https://ui.adsabs.harvard.edu/abs/2014MNRAS.445.1731F} {445, 1731}

\bibitem[\protect\citeauthoryear{{Goldreich} \& {Sari}}{{Goldreich} \&
  {Sari}}{2003}]{gs03}
{Goldreich} P.,  {Sari} R.,  2003, \mn@doi [\apj] {10.1086/346202}, \href
  {http://ui.adsabs.harvard.edu/abs/2003ApJ...585.1024G} {585, 1024}

\bibitem[\protect\citeauthoryear{{Goldreich} \& {Tremaine}}{{Goldreich} \&
  {Tremaine}}{1980}]{gt80}
{Goldreich} P.,  {Tremaine} S.,  1980, \mn@doi [\apj] {10.1086/158356}, \href
  {http://ui.adsabs.harvard.edu/abs/1980ApJ...241..425G} {241, 425}

\bibitem[\protect\citeauthoryear{{Goldreich} \& {Tremaine}}{{Goldreich} \&
  {Tremaine}}{1981}]{gt81}
{Goldreich} P.,  {Tremaine} S.,  1981, \mn@doi [\apj] {10.1086/158671}, \href
  {http://ui.adsabs.harvard.edu/abs/1981ApJ...243.1062G} {243, 1062}

\bibitem[\protect\citeauthoryear{{Hahn}}{{Hahn}}{2003}]{hahn03}
{Hahn} J.~M.,  2003, \mn@doi [\apj] {10.1086/377195}, \href
  {https://ui.adsabs.harvard.edu/abs/2003ApJ...595..531H} {595, 531}

\bibitem[\protect\citeauthoryear{{Johns-Krull} et~al.,}{{Johns-Krull}
  et~al.}{2016}]{johnskrull16}
{Johns-Krull} C.~M.,  et~al., 2016, \mn@doi [\apj]
  {10.3847/0004-637X/826/2/206}, \href
  {http://ui.adsabs.harvard.edu/abs/2016ApJ...826..206J} {826, 206}

\bibitem[\protect\citeauthoryear{{Kley} \& {Dirksen}}{{Kley} \&
  {Dirksen}}{2006}]{kd06}
{Kley} W.,  {Dirksen} G.,  2006, \mn@doi [\aap] {10.1051/0004-6361:20053914},
  \href {http://ui.adsabs.harvard.edu/abs/2006A%26A...447..369K} {447, 369}

\bibitem[\protect\citeauthoryear{{Lai}}{{Lai}}{2014}]{lai14}
{Lai} D.,  2014, \mn@doi [\mnras] {10.1093/mnras/stu485}, \href
  {http://ui.adsabs.harvard.edu/abs/2014MNRAS.440.3532L} {440, 3532}

\bibitem[\protect\citeauthoryear{{Lee} \& {Peale}}{{Lee} \&
  {Peale}}{2003}]{lp03}
{Lee} M.~H.,  {Peale} S.~J.,  2003, \mn@doi [\apj] {10.1086/375857}, \href
  {https://ui.adsabs.harvard.edu/abs/2003ApJ...592.1201L} {592, 1201}

\bibitem[\protect\citeauthoryear{{Lee}, {Dempsey}  \& {Lithwick}}{{Lee}
  et~al.}{2019}]{lee19}
{Lee} W.-K.,  {Dempsey} A.~M.,   {Lithwick} Y.,  2019, \mn@doi [\apj]
  {10.3847/1538-4357/ab010c}, \href
  {https://ui.adsabs.harvard.edu/abs/2019ApJ...872..184L} {872, 184}

\bibitem[\protect\citeauthoryear{{Lubow}}{{Lubow}}{1991}]{lubow1991a}
{Lubow} S.~H.,  1991, \mn@doi [\apj] {10.1086/170647}, \href
  {https://ui.adsabs.harvard.edu/abs/1991ApJ...381..259L} {381, 259}

\bibitem[\protect\citeauthoryear{{Lubow} \& {Ogilvie}}{{Lubow} \&
  {Ogilvie}}{2000}]{lo00}
{Lubow} S.~H.,  {Ogilvie} G.~I.,  2000, \mn@doi [\apj] {10.1086/309101}, \href
  {http://ui.adsabs.harvard.edu/abs/2000ApJ...538..326L} {538, 326}

\bibitem[\protect\citeauthoryear{{Lubow} \& {Ogilvie}}{{Lubow} \&
  {Ogilvie}}{2001}]{lo01}
{Lubow} S.~H.,  {Ogilvie} G.~I.,  2001, \mn@doi [\apj] {10.1086/322493}, \href
  {http://ui.adsabs.harvard.edu/abs/2001ApJ...560..997L} {560, 997}

\bibitem[\protect\citeauthoryear{{Muley}, {Fung}  \& {van der Marel}}{{Muley}
  et~al.}{2019}]{muley19}
{Muley} D.,  {Fung} J.,   {van der Marel} N.,  2019, \mn@doi [\apjl]
  {10.3847/2041-8213/ab24d0}, \href
  {https://ui.adsabs.harvard.edu/abs/2019ApJ...879L...2M} {879, L2}

\bibitem[\protect\citeauthoryear{{Murray} \& {Dermott}}{{Murray} \&
  {Dermott}}{1999}]{md99}
{Murray} C.~D.,  {Dermott} S.~F.,  1999, {Solar system dynamics}.
Cambridge University Press

\bibitem[\protect\citeauthoryear{{Mutter}, {Pierens}  \& {Nelson}}{{Mutter}
  et~al.}{2017}]{mutter17}
{Mutter} M.~M.,  {Pierens} A.,   {Nelson} R.~P.,  2017, \mn@doi [\mnras]
  {10.1093/mnras/stw2768}, \href
  {https://ui.adsabs.harvard.edu/abs/2017MNRAS.465.4735M} {465, 4735}

\bibitem[\protect\citeauthoryear{{Ogilvie}}{{Ogilvie}}{1999}]{ogilvie99}
{Ogilvie} G.~I.,  1999, \mn@doi [\mnras] {10.1046/j.1365-8711.1999.02340.x},
  \href {https://ui.adsabs.harvard.edu/abs/1999MNRAS.304..557O} {304, 557}

\bibitem[\protect\citeauthoryear{{Ogilvie}}{{Ogilvie}}{2001}]{ogilvie01}
{Ogilvie} G.~I.,  2001, \mn@doi [\mnras] {10.1046/j.1365-8711.2001.04416.x},
  \href {http://ui.adsabs.harvard.edu/abs/2001MNRAS.325..231O} {325, 231}

\bibitem[\protect\citeauthoryear{{Ogilvie}}{{Ogilvie}}{2006}]{ogilvie06}
{Ogilvie} G.~I.,  2006, \mn@doi [\mnras] {10.1111/j.1365-2966.2005.09776.x},
  \href {https://ui.adsabs.harvard.edu/abs/2006MNRAS.365..977O} {365, 977}

\bibitem[\protect\citeauthoryear{{Ogilvie}}{{Ogilvie}}{2007}]{ogilvie07}
{Ogilvie} G.~I.,  2007, \mn@doi [\mnras] {10.1111/j.1365-2966.2006.11141.x},
  \href {http://ui.adsabs.harvard.edu/abs/2007MNRAS.374..131O} {374, 131}

\bibitem[\protect\citeauthoryear{{Ogilvie}}{{Ogilvie}}{2018}]{ogilvie18}
{Ogilvie} G.~I.,  2018, \mn@doi [\mnras] {10.1093/mnras/sty588}, \href
  {https://ui.adsabs.harvard.edu/abs/2018MNRAS.477.1744O} {477, 1744}

\bibitem[\protect\citeauthoryear{{Ogilvie} \& {Latter}}{{Ogilvie} \&
  {Latter}}{2013}]{ol13}
{Ogilvie} G.~I.,  {Latter} H.~N.,  2013, \mn@doi [\mnras]
  {10.1093/mnras/stt916}, \href
  {https://ui.adsabs.harvard.edu/abs/2013MNRAS.433.2403O} {433, 2403}

\bibitem[\protect\citeauthoryear{{Ogilvie} \& {Lubow}}{{Ogilvie} \&
  {Lubow}}{2003}]{ol03}
{Ogilvie} G.~I.,  {Lubow} S.~H.,  2003, \mn@doi [\apj] {10.1086/368178}, \href
  {http://ui.adsabs.harvard.edu/abs/2003ApJ...587..398O} {587, 398}

\bibitem[\protect\citeauthoryear{{Papaloizou} \& {Lin}}{{Papaloizou} \&
  {Lin}}{1995}]{pl95}
{Papaloizou} J.~C.~B.,  {Lin} D.~N.~C.,  1995, \mn@doi [\apj] {10.1086/175127},
  \href {http://ui.adsabs.harvard.edu/abs/1995ApJ...438..841P} {438, 841}

\bibitem[\protect\citeauthoryear{{Papaloizou} \& {Pringle}}{{Papaloizou} \&
  {Pringle}}{1983}]{pp83}
{Papaloizou} J.~C.~B.,  {Pringle} J.~E.,  1983, \mn@doi [\mnras]
  {10.1093/mnras/202.4.1181}, \href
  {https://ui.adsabs.harvard.edu/abs/1983MNRAS.202.1181P} {202, 1181}

\bibitem[\protect\citeauthoryear{{Papaloizou}, {Nelson}  \&
  {Masset}}{{Papaloizou} et~al.}{2001}]{papaloizou01}
{Papaloizou} J.~C.~B.,  {Nelson} R.~P.,   {Masset} F.,  2001, \mn@doi [\aap]
  {10.1051/0004-6361:20000011}, \href
  {http://ui.adsabs.harvard.edu/abs/2001A%26A...366..263P} {366, 263}

\bibitem[\protect\citeauthoryear{{Petrovich}, {Wu}  \& {Ali-Dib}}{{Petrovich}
  et~al.}{2019}]{petrovich19}
{Petrovich} C.,  {Wu} Y.,   {Ali-Dib} M.,  2019, \mn@doi [\aj]
  {10.3847/1538-3881/aaeed9}, \href
  {http://ui.adsabs.harvard.edu/abs/2019AJ....157....5P} {157, 5}

\bibitem[\protect\citeauthoryear{{Ragusa}, {Rosotti}, {Teyssandier}, {Booth},
  {Clarke}  \& {Lodato}}{{Ragusa} et~al.}{2018}]{ragusa18}
{Ragusa} E.,  {Rosotti} G.,  {Teyssandier} J.,  {Booth} R.,  {Clarke} C.~J.,
  {Lodato} G.,  2018, \mn@doi [\mnras] {10.1093/mnras/stx3094}, \href
  {http://ui.adsabs.harvard.edu/abs/2018MNRAS.474.4460R} {474, 4460}

\bibitem[\protect\citeauthoryear{{Reg{\'a}ly}, {S{\'a}ndor}, {Dullemond}  \&
  {van Boekel}}{{Reg{\'a}ly} et~al.}{2010}]{regaly10}
{Reg{\'a}ly} Z.,  {S{\'a}ndor} Z.,  {Dullemond} C.~P.,   {van Boekel} R.,
  2010, \mn@doi [\aap] {10.1051/0004-6361/201014427}, \href
  {http://ui.adsabs.harvard.edu/abs/2010A%26A...523A..69R} {523, A69}

\bibitem[\protect\citeauthoryear{{Rosotti}, {Booth}, {Clarke}, {Teyssandier},
  {Facchini}  \& {Mustill}}{{Rosotti} et~al.}{2017}]{rosotti}
{Rosotti} G.~P.,  {Booth} R.~A.,  {Clarke} C.~J.,  {Teyssandier} J.,
  {Facchini} S.,   {Mustill} A.~J.,  2017, \mn@doi [\mnras]
  {10.1093/mnrasl/slw184}, \href
  {https://ui.adsabs.harvard.edu/abs/2017MNRAS.464L.114R} {464, L114}

\bibitem[\protect\citeauthoryear{{Sefilian} \& {Rafikov}}{{Sefilian} \&
  {Rafikov}}{2019}]{sr19}
{Sefilian} A.~A.,  {Rafikov} R.~R.,  2019, arXiv e-prints, \href
  {https://ui.adsabs.harvard.edu/abs/2019arXiv190407592S} {p. arXiv:1904.07592}

\bibitem[\protect\citeauthoryear{{Teyssandier} \& {Ogilvie}}{{Teyssandier} \&
  {Ogilvie}}{2016}]{to16}
{Teyssandier} J.,  {Ogilvie} G.~I.,  2016, \mn@doi [\mnras]
  {10.1093/mnras/stw521}, \href
  {http://ui.adsabs.harvard.edu/abs/2016MNRAS.458.3221T} {458, 3221}

\bibitem[\protect\citeauthoryear{{Teyssandier} \& {Ogilvie}}{{Teyssandier} \&
  {Ogilvie}}{2017}]{to17}
{Teyssandier} J.,  {Ogilvie} G.~I.,  2017, \mn@doi [\mnras]
  {10.1093/mnras/stx426}, \href
  {http://ui.adsabs.harvard.edu/abs/2017MNRAS.467.4577T} {467, 4577}

\bibitem[\protect\citeauthoryear{{Touma}}{{Touma}}{2002}]{touma02}
{Touma} J.~R.,  2002, \mn@doi [\mnras] {10.1046/j.1365-8711.2002.05437.x},
  \href {https://ui.adsabs.harvard.edu/abs/2002MNRAS.333..583T} {333, 583}

\bibitem[\protect\citeauthoryear{{Tremaine}}{{Tremaine}}{2001}]{tremaine01}
{Tremaine} S.,  2001, \mn@doi [\aj] {10.1086/319398}, \href
  {https://ui.adsabs.harvard.edu/abs/2001AJ....121.1776T} {121, 1776}

\bibitem[\protect\citeauthoryear{{Zanazzi} \& {Lai}}{{Zanazzi} \&
  {Lai}}{2018a}]{zl18a}
{Zanazzi} J.~J.,  {Lai} D.,  2018a, \mn@doi [\mnras] {10.1093/mnras/stx2375},
  \href {https://ui.adsabs.harvard.edu/abs/2018MNRAS.473..603Z} {473, 603}

\bibitem[\protect\citeauthoryear{{Zanazzi} \& {Lai}}{{Zanazzi} \&
  {Lai}}{2018b}]{zl18b}
{Zanazzi} J.~J.,  {Lai} D.,  2018b, \mn@doi [\mnras] {10.1093/mnras/sty951},
  \href {https://ui.adsabs.harvard.edu/abs/2018MNRAS.477.5207Z} {477, 5207}

\bibitem[\protect\citeauthoryear{{Zanazzi} \& {Lai}}{{Zanazzi} \&
  {Lai}}{2018c}]{zl18c}
{Zanazzi} J.~J.,  {Lai} D.,  2018c, \mn@doi [\mnras] {10.1093/mnras/sty1075},
  \href {https://ui.adsabs.harvard.edu/abs/2018MNRAS.478..835Z} {478, 835}

\bibitem[\protect\citeauthoryear{{Zhang}, {Hamilton}  \& {Matsumura}}{{Zhang}
  et~al.}{2013}]{zhang13}
{Zhang} K.,  {Hamilton} D.~P.,   {Matsumura} S.,  2013, \mn@doi [\apj]
  {10.1088/0004-637X/778/1/6}, \href
  {https://ui.adsabs.harvard.edu/abs/2013ApJ...778....6Z} {778, 6}

\makeatother
\end{thebibliography}

\appendix

\section{Precession and damping rates: expressions for dimensionless coefficients}
\label{app:hatw}
In order to compute all the dimensionless coefficients $\hat{\omega}$ in Section \ref{sec:toy}, it is useful to introduce $x=r/\rin$. Unless mentioned otherwise, all the integrals in this Appendix are carried out from 1 to $x_{\rm out}=\rout/\rin$.

\subsection{Pressure and viscosity}
\label{app:hatwpv}
For the pressure term, the dimensionless coefficient is
\begin{equation}
\hat{\omega}_{\rm d,pr} =  \frac{\int g(x)~\id x}{\int f^2(x)~ s(x)~ x^{3/2}~\id x},
\end{equation}
where
\begin{align}
g(x) &= -\left( \dd{f}{x}\right)^2~ s^2(x)~ x^2 + x\left[f^2(x)~\dd{s}{x}-f(x)~\dd{f}{x}~s(x) \right] \nonumber \\
&+ 2f^2(x)~s(x).
\end{align}
For the viscosity term, we have
\begin{align}
\hat{\omega}_{\rm d,visc} =\frac{\int \left( \dd{f}{x}\right)^2~ s^2(x)~ x^2~\id x}{\int f^2(x)~ s(x)~ x^{3/2}~\id x}.
\end{align}

\subsection{Self-gravity}
\label{app:hatwsg}
The precession rate of a self-gravitating disc can be expressed as:
\begin{equation}
\label{eq:wsgapp}
\omega_{\rm d, sg}=\frac{I_{\rm d,sg}}{2A},
\end{equation}
with
\begin{align}
\label{eq:intddapp}
I_{\rm d,sg} & = \iint \frac{1}{4}G\Sigma(r)\Sigma (r') \bigg\{ \left[ K_1 (r,r')+K_2 (r,r') \right]|E(r)-E(r')|^2\nonumber\\
& + \left[ K_1 (r,r')-K_2 (r,r') \right]|E(r)+E(r')|^2\bigg\} 2\pi r \,\id r \, 2\pi r' \,\id r'.
\end{align}
The coefficents $K_1$ and $K_2$ are smoothed kernels given by
\begin{equation}
\label{eq:K}
K_m(r,r')=\frac{rr'}{4\pi}\int_{0}^{2\pi}\frac{\cos m\theta \id \theta}{\left[r^2+r'^2-2rr'\cos\theta+s^2rr'\right]^{3/2}}.
\end{equation}
Here $s$ is a dimensionless smoothing parameter that prevents $K$ to diverge when $r=r'$. This smoothing needs to be introduced when computing the disc's self-gravity, but is not required when computing planet-disc interactions. We can write $K$ in terms of the usual Laplace coefficients of celestial mechanics,
\begin{equation}
b_{3/2}^m(\beta)=\frac{1}{\pi}\int_{0}^{2\pi}\frac{\cos m\theta \id \theta}{\left[1-2\beta\cos\theta+\beta^2\right]^{3/2}},
\end{equation} 
as
\begin{equation}
K_m(r,r')=\frac{\beta^{3/2}}{4(rr')^{1/2}}b_{3/2}^m(\beta),
\end{equation}
where $\beta$ is the solution of 
\begin{equation}
\frac{1+\beta^2}{\beta}=\frac{r^2+r'^2}{rr'}+s^2,
\end{equation}
such as $\beta<1$. We take the smoothing length $s$ to be the disc aspect ratio $h$. %Note that for planet-disc interactions, we do not need to introduce this smoothing length because the integrals do not diverge. 

It is possible to write the precession rate given by equation (\ref{eq:wsgapp}) as
\begin{equation}
\omega_{\rm d,sg}=\frac{\Sigma_0 \rin^2}{\Ms}\Oin~\hat{\omega}_{\rm d,sg},
\end{equation}
where 
\begin{align}
&\hat{\omega}_{\rm d,sg} = \frac{I_1}{{2\int f^2(x)~s(x)~x^{3/2}~\id x}},
\end{align}
with
%\begin{align}
%I_1&=\iint_1^{x_{\rm out}}s(x)s(x')\left[\left(F_1(x,x')+F_2(x,x')\right)(f(x)-f(x'))^2 \right. \nonumber \\
%&\left.  +\left(F_1(x,x')-F_2(x,x')\right)(f(x)+f(x'))^2 \right] \id x\id x'.
%\end{align}
\begin{align}
I_{\rm 1} & = \iint s(x)~s(x')~ \bigg\{ \left[ F_1 (x,x')+F_2 (x,x') \right]|f(x)-f(x')|^2\nonumber\\
& + \left[ F_1 (x,x')-F_2 (x,x') \right]|f(x)+f(x')|^2\bigg\}~ x~\id x~ x'~\id x'.
\end{align}
We recall that $x=r/\rin$ is a dimensionless variable, and we have introduced
\begin{align}
F_m(x,x')&= xx'\int_{0}^{2\pi}\frac{\cos m\theta \id \theta}{\left[x^2+x'^2-2xx'\cos\theta+s^2xx'\right]^{3/2}}.
\end{align}
We note that computing $I_1$ requires to numerically compute three integrals: a double integral over the disc radius, and the $F$ coefficients from 0 to $2\pi$. This can be numerically time-consuming. The computation time can be significantly improved by noting that $F_m(x,x')=(\pi\beta^{3/2}/(xx')^{1/2})b_{3/2}^m(\beta)$, where $\beta$ is once again the lower-than-one solution of 
\begin{equation}
\frac{1+\beta^2}{\beta}=\frac{x^2+x'^2}{xx'}+s^2.
\end{equation} 
The Laplace coefficients can be expressed in terms of complete elliptical integrals of the first and second kind, $K$ and $E$ \citep[see, e.g.][]{BC61}:
\begin{align}
b_{3/2}^1(\beta)=\frac{4}{\pi}\frac{1}{(1-\beta^2)^2}\left[(1+\beta^2)E(\beta)-(1-\beta^2)K(\beta) \right],
\end{align}
\begin{align}
b_{3/2}^2(\beta)=\frac{4}{\pi}\frac{1}{\alpha^2(1-\beta^2)^2}&\left[2(\beta^4-\beta^2+1)E(\beta)\right. \nonumber \\
&\left. -(\beta^4-3\beta^2-2)K(\beta) \right].
\end{align}
Because complete elliptical integrals are well tabulated in many programming languages, we found that re-writing the $F$ functions in such way significantly improve the time it takes to compute these integrals.

\subsection{Planet-disc coupling}
\label{app:hatwpd}
The dimensionless coefficients that enter the planet-disc secular coupling equations are given by:
\begin{align}
\hat{\omega}_{\rm d,p} &= \frac{\int f^2(x)~ s(x)~ x^{-2}~\id x}{\int f^2(x)~ s(x)~ x^{3/2}~\id x},\\
\hat{\nu}_{\rm d,p} &= \frac{\int f(x)~ s(x)~ x^{-3}~\id x}{\int f^2(x)~ s(x)~ x^{3/2}~\id x},\\
\hat{\omega}_{\rm p,d} &= \int s(x)~x^{-2}~\id x,\\
\hat{\nu}_{\rm p,d} &= \int f(x)~s(x)~x^{-3}~\id x,.
\end{align}

\section{Planet-planet interactions}
\label{app:pp}
Consider two planets of mass $m_i$, semi-major axis $a_i$, eccentricity $e_i$ and mean anomaly $n_i$ for $i={1,2}$ with $a_1<a_2$. At octupole order, the complex eccentricity equations are given by \citep[see, e.g.,][]{lp03}:
\begin{equation}
\dd{E_1}{t}=\im \omega_{11}\frac{X_1^{1/2}}{X_2^{3/2}}E_1 - \im \omega_{12}\frac{X_1^{1/2}}{2X_2^{5/2}}\left[\frac{3}{2}E_1^2E_2^*+Y_1E_2\right],
\end{equation}
and
\begin{equation}
\dd{E_2}{t}=\im \omega_{22}\frac{V_1}{X_2^{2}}E_2 - \im \omega_{21}\frac{Z_1}{2X_2^3}\left[5E_2^2E_1^*+Y_2E_1\right],
\end{equation}
where
\begin{equation*}
X_i=1-e_i^2, \quad Y_i=2+3e_i^2, \quad Z_i=1+\frac{3}{4}e_i^2, \quad V_i=1+\frac{3}{2}e_i^2.
\end{equation*}
The frequencies are given by
\begin{align}
\omega_{11}&=\frac{3}{4}\frac{m_2}{\Ms}\left(\frac{a_1}{a_2}\right)^3n_1, \qquad \omega_{12}=\frac{15}{16}\frac{m_2}{\Ms}\left(\frac{a_1}{a_2}\right)^4n_1, \nonumber \\
\omega_{22}&=\frac{3}{4}\frac{m_1}{\Ms}\left(\frac{a_1}{a_2}\right)^2n_2, \qquad \omega_{21}=\frac{15}{16}\frac{m_1}{\Ms}\left(\frac{a_1}{a_2}\right)^3n_2.
\end{align}

%\section{Non-linear planet-disc interactions}
%Using the formalism developed by \citet{ogilvie07}, one can extract the equations coupling the eccentricity dynamics of a planet-disc system at order $e^3$. For the planet we have:
%\begin{align}
%\label{eq:edp_nl}
%\Mp\ap^2\Op\dd{\Ep}{t}&=\int\im \frac{G\Mp\Sigma}{\ap}  \bigg\{ 2f_2\Ep+f_{10}E \nonumber\\
%&+ \left[\left(4f_6-f_2\right)\Ep+\left(2f_{12}-\frac{1}{2}f_{10}\right)E \right]|\Ep|^2 \nonumber\\
%&+\left[ 2f_5\Ep+f_{11}E \right]|E|^2 \nonumber\\
%&+f_{12} E^*\Ep^2 + 2f_{17}E^2\Ep^* \bigg\}\twopi
%\end{align}
%and for a ring of the disc we have
%\begin{align}
%\label{eq:epd_nl}
%\Sigma r^2 \Omega \dd{E}{t}&=\im \frac{G\Mp\Sigma}{\ap}  \bigg\{ 2f_2E+f_{10}\Ep \nonumber\\
%&+ \left[\left(4f_4-f_2\right)E+\left(2f_{11}-\frac{1}{2}f_{10}\right)\Ep \right]|E|^2 \nonumber\\
%&+\left[ 2f_5E+f_{12}\Ep \right]|\Ep|^2 \nonumber\\
%&+f_{11}E^2\Ep^* + 2f_{17}E^*\Ep^2 \bigg\} 
%\end{align}
%where the coefficients $f_i$ can be found in appendix B of \citet{md99}. Note that $f_2=(1/8)(\ap/r)b_{3/2}^1$ and $f_{10}=-(1/4)(\ap/r)b_{3/2}^2$, and one can recover  recover equations (\ref{eq:epd}) and (\ref{eq:edp}) at linear order in eccentricity.
%
%At the leading non-linear order, the AMD of the planet and disc read
%\begin{align}
%A_p&=\frac{1}{2}\Mp\ap^2\Op\left(|\Ep|^2+\frac{1}{4}|\Ep|^4\right), \\
%A_d&=\int\frac{1}{2}\Sigma r^2\Omega\left(|E|^2+\frac{1}{4}|E|^4 \right)\twopi.
%\end{align}
%One can show that in this case, equations (\ref{eq:edp_nl}) and (\ref{eq:epd_nl}) satisfy the conservation of total AMD.

\label{lastpage}
\end{document}